\documentclass{article}

\PassOptionsToPackage{numbers, compress, sort}{natbib}
\PassOptionsToPackage{capitalise,noabbrev}{cleveref} 

\PassOptionsToPackage{table,dvipsnames}{xcolor} 

\usepackage[preprint]{neurips_2025} 

\usepackage[utf8]{inputenc} 
\usepackage[T1]{fontenc}    

\usepackage{url}            
\usepackage{booktabs}       
\usepackage{amsfonts}       
\usepackage{nicefrac}       
\usepackage{microtype}      
\usepackage{graphicx}       
\usepackage{amsmath, amssymb, amsthm} 
\usepackage{bm}             
\usepackage{siunitx}        
\usepackage{multirow}       
\usepackage{makecell}       
\usepackage{caption, subcaption} 
\usepackage{enumitem}       

\usepackage{algorithm}      
\usepackage{algpseudocode}  

\usepackage{tikz}           
\usepackage{pgfplots}       
\pgfplotsset{compat=1.18}   
\usepackage{colortbl}       
\usepackage[most]{tcolorbox}
\usepackage{pgf-pie}        

\usepackage{wrapfig}        
\usepackage{picinpar}       
\usepackage{cutwin}         
\usepackage{tabularx}       
\usepackage{mathrsfs}       
\usepackage{newfloat}       
\usepackage{listings}       
\usepackage{pifont}         
\usepackage{supertabular}   
\usepackage{array}          

\usepackage{appendix}       
\usepackage{stfloats}       
\usepackage{afterpage}      

\usepackage{hyperref}       
\usepackage{cleveref}       

\newtheorem{definition}{Definition}[section]
\newtheorem{fact}{Fact}[section]

\DeclareMathOperator*{\argmin}{arg\,min}

\crefname{section}{Section}{Sections}
\crefname{appendix}{Appendix}{Appendices}
\crefname{figure}{Figure}{Figures}
\crefname{table}{Table}{Tables}
\crefname{equation}{Eq.}{Eqs.}
\crefname{algorithm}{Algorithm}{Algorithms}
\crefname{fact}{Fact}{Facts}
\crefname{definition}{Definition}{Definitions}
\crefname{lemma}{Lemma}{Lemmas}
\crefname{proposition}{Proposition}{Propositions}

\definecolor{tableheadcolor}{gray}{0.8}
\definecolor{tablegray}{gray}{0.9} 
\definecolor{ourscolor}{HTML}{E0EFFF} 
\definecolor{ourscolor_orange}{HTML}{FFE5D0} 
\definecolor{myhighlightcolor}{gray}{0.9} 

\newtcolorbox{mygraybox}{
  colback=gray!10,
  colframe=black,
  boxrule=1.5pt,
  arc=4pt
}

\usetikzlibrary{patterns} 
\newcommand{\partialvrule}[1]{
  \tikz[overlay]{\draw (0,-1.1ex) -- (0,2.5ex);}%
}

\title{Dual-Priv Pruning : Efficient Differential Private Fine-Tuning in Multimodal Large Language Models}

\author{%
  Qianshan Wei\textsuperscript{1}\thanks{These authors contributed equally to this work.}\hspace{0.5em}, 
  Jiaqi Li\textsuperscript{1}\footnotemark[1]\hspace{0.5em}, 
  Zihan You\textsuperscript{2}\footnotemark[1]\hspace{0.5em}, 
  Yi Zhan\textsuperscript{3}\hspace{0.5em},
  Kecen Li\textsuperscript{4}\hspace{0.5em},
  Jialin Wu\textsuperscript{5}\hspace{0.5em},
  Xinfeng Li\textsuperscript{5,6} \\  
  \textbf{Hengjun
  Liu}\textsuperscript{7}\hspace{0.5em},
  \textbf{Yi Yu}\textsuperscript{6}\hspace{0.5em},
  \textbf{Bin Cao}\textsuperscript{4}\hspace{0.5em},
  \textbf{Yiwen Xu}\textsuperscript{8}\hspace{0.5em},
  \textbf{Yang Liu}\textsuperscript{6}\hspace{0.5em},
  \textbf{Guilin Qi}\textsuperscript{1}\thanks{Corresponding author. }\\[2ex] 
  \noindent\textsuperscript{1}School of Cyber Science and Engineering, Southeast University, Nanjing, China \\
  \noindent\textsuperscript{2}School of Automation, Southeast University, Nanjing, China \\
  \noindent\textsuperscript{3}School of Computer Science, Peking University, Beijing, China \\
  \noindent\textsuperscript{4}Institute of Automation, Chinese Academy of Sciences (CASIA), Beijing, China \\
  \noindent\textsuperscript{5}Zhejiang University, Hangzhou, China \\
  \noindent\textsuperscript{6}Nanyang Technological University, Singapore \\
  \noindent\textsuperscript{7}Electrical Engineering and Information Technology, Technische Universität Chemnitz, Chemnitz, Germany \\ 
  \noindent\textsuperscript{8}University of California, Los Angeles (UCLA), Los Angeles, CA, USA
}

\begin{document}

\maketitle

\begin{abstract}
  Differential Privacy (DP) is a widely adopted technique, valued for its effectiveness in protecting the privacy of task-specific datasets, making it a critical tool for large language models. However, its effectiveness in Multimodal Large Language Models (MLLMs) remains uncertain. Applying Differential Privacy (DP) inherently introduces substantial computation overhead, a concern particularly relevant for MLLMs which process extensive textual and visual data. Furthermore, a critical challenge of DP is that the injected noise, necessary for privacy, scales with parameter dimensionality, leading to pronounced model degradation; This trade-off between privacy and utility complicates the application of Differential Privacy (DP) to complex architectures like MLLMs. To address these, we propose \textbf{Dual-Priv Pruning}, a framework that employs two complementary pruning mechanisms for DP fine-tuning in MLLMs: (i) \textit{visual token pruning} to reduce input dimensionality by removing redundant visual information, and (ii) \textit{gradient-update pruning} during the DP optimization process. This second mechanism selectively prunes parameter updates based on the magnitude of noisy gradients, aiming to mitigate noise impact and improve utility. Experiments demonstrate that our approach achieves competitive results with minimal performance degradation. In terms of computational efficiency, our approach consistently utilizes less memory than standard DP-SGD. While requiring only 1.74\% more memory than zeroth-order methods which suffer from severe performance issues on A100 GPUs, our method demonstrates leading memory efficiency on H20 GPUs. To the best of our knowledge, we are the first to explore DP fine-tuning in MLLMs. Our code is coming soon.
\end{abstract}

\section{Introduction}
\vspace{-2mm}
 Large Language Models (LLMs)~\cite{zhang2022optopenpretrainedtransformer,radford2019language,touvron2023llamaopenefficientfoundation} have showcased remarkable proficiency in natural language processing, driving their widespread adoption in downstream tasks~\cite{ziegler2019fine}, and Multimodal Large Language Models (MLLMs)~\cite{liu2023visual,wang2024qwen2,abdin2024phi}extend the power of LLMs by integrating text and visual data, opening up possibilities for applications that require understanding across different modalities. However, both models are easy to risk leaking sensitive information during training~\cite{das2025security,mesko2023impact}. Differential Privacy~\cite{dwork2006differential} (DP) , the technology for providing privacy guarantees that limit the ability to infer whether a data point was used in the training process of a model by observing its output. This technology is typically achieved by injecting noise during training processes, limiting the discernible impact of single data point. The degree of privacy guarantee is tuned using a privacy budget ($\epsilon$), where stronger privacy guarantee (lower $\epsilon$) generally comes at the cost of adding more noise and degrading model performance. The inherent trade-off between privacy and utility presents a significant challenge, particularly when applying DP to large and complex models like LLMs, since the necessary noise often scales with parameter dimensionality. Prior works~\cite{yu2021differentially,li2021large,liu2024differentially,goel2025differentially} have shown that LLMs with hundreds of millions of parameters can be effectively and efficiently fine-tuned to yield models with high performance under modest privacy leakage.

However, it remains unclear whether such conclusions of LLMs are transferable to MLLMs. Similar to unimodal models, DP also face challenges under MLLMs. The first is \textbf{computation consumption}. This challenge is exacerbated in MLLMs, which rely on a large number of visual tokens ( e.g., 197 tokens per image in CLIP-ViT~\cite{radford2021learning} or hundreds in LLaVA~\cite{liu2023visual} ) to represent detailed visual information, significantly increasing computation demands. Recent work~\cite{li2021large}introduced ``ghost clipping'' to address the computation overhead of DP-SGD in LLMs. While ghost clipping reduces computation overhead by leveraging the sequential structure of text, its reliance on this sequentiality renders it unsuitable for image features, as MLLMs process these features as non-sequential data within their multimodal components. Zeroth-order methods (e.g., DP-ZO~\cite{tang2024private}) also aim to reduce computation overhead by avoiding explicit gradient calculations. However, these methods introduce severe convergence issues. For instance, DP-ZO required more training steps (75k vs 200) than standard DP-SGD to achieve comparable performance on SQuAD~\cite{tang2024private}, making this gradient-free approach prohibitively slow for practical MLLM training. Another challenge is \textbf{model degradation}. Differential privacy introduces noise to safeguard data privacy, but this noise perturbs the gradient signals during training, leading to performance degradation. In MLLMs, DP noise scales with parameter dimensionality, overwhelming gradient signals in high-dimensional layers and necessitating more iterations to stabilize optimization, as noted in foundational work on DP-SGD~\cite{abadi2016deep}.

To tackle these challenges, we introduce \textbf{Dual-Priv Pruning}, a novel DP finetuning approach tailored for MLLMs. Our approach integrates two complementary
pruning mechanisms designed to work in concert, addressing these issues from both the input representation and the optimization process. The first key pruning mechanism focuses on optimizing the visual input stream prior to training: it employs an attention-based mechanism to identify and prune redundant visual tokens, thereby substantially reducing the input dimensionality and subsequent computational demands. The less critical visual information pruned in this manner is then fused into some compact contextual representations, to which a calibrated heuristic noise is added. This step aims to preserve essential global context while further alleviating the processing load for the differential privacy mechanism. The second core pruning mechanism refines the differential private fine-tuning process itself. While adhering to the standard DP-SGD framework for rigorous noise addition to guarantee privacy, Dual-Priv Pruning introduces a  \textit{gradient-update pruning} technique. This technique analyzes the noisy gradients resulting from DP noise injection. It then selectively applies these gradients for parameter updates only to those blocks where the underlying signal is deemed sufficiently strong and reliable to overcome the obfuscating effect of the DP noise, thereby preserving model utility and stabilizing training.
Dual-Priv Pruning offers a robust solution. As the first work to explore DP finetuning specifically tailored for MLLMs, our method achieves a superior privacy-utility trade-off and enhanced computational efficiency, delivering competitive performance even under stringent privacy budgets.

We summarize our main contributions as follows:
\textbf{(1)} We pioneer the integration of DP into the domain of MLLMs, addressing a critical research gap in privacy-preserving multimodal learning.
\textbf{(2)} We introduce a novel privacy-aware visual pruning mechanism that significantly reduces computational overhead by optimizing visual inputs, thereby creating more favorable conditions for subsequent DP fine-tuning.
\textbf{(3)} We propose an DP-compatible gradient-update pruning strategy that intelligently applies noisy gradients to mitigate the adverse effects of DP noise on model performance, thereby enhancing utility while maintaining strong privacy guarantees.
\textbf{(4)} Extensive experiments demonstrate that our Dual-Priv Pruning achieves robust privacy protection, substantial memory reduction, and competitive performance on diverse vision-language tasks, even under stringent privacy budgets.
    
    


\vspace{-2mm}
\section{Related Work}
\vspace{-3mm}
\noindent \textbf{Differential Privacy (DP)}~\cite{dwork2006differential} ensures privacy guarantees by limiting the ability to infer whether a data point was used in the training process of a model, making it a cornerstone for privacy-preserving learning. In the area of computer vision, \cite{tang2023differentially} developed DP methods for image classification by adding noisy priors, achieving strong privacy-utility trade-offs, and \cite{luo2024differentially} applied DP to video  recognition, enforcing video-level differential privacy through clip-based classification models. In natural language processing, \cite{mcmahan2017learning} trained recurrent language models with DP, reducing risks of data memorization. For LLMs, \cite{li2024fine} demonstrated DP fine-tuning but noted challenges with utility degradation due to noise, while \cite{kerrigan2020differentially} showed that public pre-training followed by private fine-tuning can alleviate some performance losses. Memory-efficient techniques, such as ``ghost-clipping''~\cite{li2021large}, optimize DP-SGD for LLMs but rely on text-specific assumptions, limiting their applicability to multimodal settings. Zeroth-order optimization~\cite{tang2024private} offers an alternative for LLMs by avoiding gradient instantiation, but it suffers from too long training times. Other efforts to improve DP include manipulating gradients, such as GIP~\cite{yangimproving} that perturbed individual gradient indices, though its privacy analysis clarity was questioned; In contrast, our gradient-update pruning operates as a post-processing step on entire noised logical parameter blocks, simplifying privacy analysis and aligning with PEFT. In multimodal learning, \cite{huang2023safeguarding} introduced DP to CLIP training, protecting vision-language data, and \cite{yu2021large} proposed low-rank reparametrization for scalable private learning, applicable to multimodal tasks. Additionally, \cite{kaissis2021end} applied DP to medical image, emphasizing privacy in sensitive domains. Despite these advances, no prior work has explored DP fine-tuning for MLLMs, which face unique challenges due to cross-modal interactions and massive length visual tokens. Existing methods, do not address the memory demands and model degradation of MLLMs, a gap that our work to addresses.

\noindent \textbf{Multimodal Large Language Models (MLLMs)} integrate visual and textual modalities to solve a wide range of tasks. Flamingo~\cite{alayrac2022flamingo} introduced a query-based cross-attention mechanism to enable multimodal interactions, while BLIP-2~\cite{li2023blip} proposed the lightweight Q-Former to enhance efficiency. InstructBLIP~\cite{dai2023instructblipgeneralpurposevisionlanguagemodels} further aligned models with user intent via instruction tuning across diverse datasets. LLaVA~\cite{liu2023visual} improved visual understanding using curated training data, while subsequent efforts such as Qwen-VL~\cite{bai2023qwen} and CogVLM~\cite{wang2024cogvlm} introduced advanced training strategies and modular visual expert systems to boost performance.
A major challenge in MLLMs is the redundancy of visual tokens, which significantly increases memory and computational costs~\cite{DBLP:conf/eccv/ChenZLBLZC24}. Recent work addresses this inefficiency: FastVLM~\cite{vasu2024fastvlm} prunes tokens based on attention scores, and VisionZip~\cite{yang2024visionzip} identifies contextual tokens that retain global semantics (e.g., background information). Visual token redundancy offers a promising avenue for DP in MLLMs. Pruning low-importance tokens reduces sensitive data exposure. We leverage this property to enable even source-level privacy protection and efficient DP fine-tuning.
\vspace{-2mm}
\section{Preliminaries}
\label{sec:preliminaries}
\vspace{-3mm}
\subsection{Differential Privacy}
\vspace{-2mm}
Differential privacy (DP)~\cite{dwork2006differential} provides a rigorous framework to safeguard sensitive data by ensuring that model outputs remain statistically indistinguishable for datasets differing by a single record. This guarantee inherently limits the ability of inferring  individual record participation, mitigating risks such as membership inference attacks~\cite{shokri2017membership}. A hallmark of DP is its robustness to post-processing: if an algorithm $\mathcal{A}$ satisfies ($\epsilon, \delta$)-DP, any function $f \circ \mathcal{A}$ preserves the same ($\epsilon, \delta$)-DP guarantee.

\begin{definition}[($\epsilon, \delta$)-Differential Privacy]
A randomized algorithm $\mathcal{A}$ is ($\epsilon, \delta$)-differentially private if, for any two neighboring datasets $\mathcal{D}$ and $\mathcal{D}'$, differing by one record, and any set of outputs $S \subseteq \text{Range}(\mathcal{A})$, the following holds:
 \label{def:dp}
\begin{equation}
\Pr[\mathcal{A}(\mathcal{D}) \in S] \leq e^{\epsilon} \cdot \Pr[\mathcal{A}(\mathcal{D}') \in S] + \delta, \label{eq:dp}
\end{equation}
\noindent where $\epsilon \geq 0$ is the privacy budget, controlling the strength of the privacy guarantee, and $\delta \in [0,1)$ is a small failure probability. 
\end{definition}

In the context of fine-tuning MLLMs, two datasets $\mathcal{D}$ and $\mathcal{D}'$ are defined as neighboring if one can be obtained from the other by adding or removing a single image-text pair. The application of DP in iterative training (introduced in \cref{def:dpsgd}),
relies on fundamental mechanisms and accounting principles. The Gaussian Mechanism (detailed in Fact~\ref{fact:gaussian_mechanism}) is employed to add noise. To manage the overall privacy loss across multiple iterations, privacy accounting techniques like Rényi Differential Privacy (RDP) (detailed in Fact~\ref{fact:composition_accounting}) are utilized. These principles are central to the DP application.

\subsubsection{Differentially Private SGD}\label{def:dpsgd} 
\vspace{-2mm}
 Differentially Private Stochastic Gradient Descent (DP-SGD)~\cite{abadi2016deep} adapts SGD to ensure the trained model parameters $\theta \in \mathbb{R}^d$ satisfy an overall $(\epsilon, \delta)$-DP guarantee with respect to $\mathcal{D}_{\text{train}}$. In each iteration $k$, for a minibatch $\xi_k$ of size $m$ sampled with probability $q = m/N$: First, per-sample gradients $g_i = \nabla_{\theta} \mathcal{L}(\theta_{k-1}, (\mathcal{I}_i, \mathcal{T}_i))$ are computed for each $i \in \xi_k$. Second, to bound sensitivity, the $L_2$ norm of each gradient $g_i$ is clipped using a threshold $C$: $\hat{g}_i = g_i / \max(1, \|g_i\|_2 / C)$. This ensures $\|\hat{g}_i\|_2 \le C$, thereby limiting the influence of any single sample and resulting in an $L_2$ sensitivity of $\Delta f = C/m$ for the subsequent average gradient (details in Appendix~\ref{app:sensitivity_derivation_content}). Third, these clipped gradients are aggregated by averaging: $\bar{g} = \frac{1}{m} \sum_{i \in \xi_k} \hat{g}_i$. Finally, calibrated Gaussian noise is added to this average gradient before updating:
\begin{equation}
\theta_k = \theta_{k-1} - \eta \cdot \left( \bar{g} + \mathcal{N}(0, \sigma^2 C^2 I_d / m^2) \right). \label{eq:dpsgd_update} 
\end{equation}
The hyperparameters $C$ (clipping norm) and $\sigma$ (noise multiplier) control the trade-off between privacy and utility. The appropriate value for $\sigma$ is determined based on the overall privacy budget $(\epsilon, \delta)$, total training steps, and sampling rate, typically using privacy accounting methods like RDP (Fact~\ref{fact:composition_accounting}).

\vspace{-2mm}
\subsection{Problem Definition: Differentially Private Fine-Tuning of MLLMs}
\label{sec:problem_definition}
\vspace{-2mm} 

Our work focuses on fine-tuning a pre-trained MLLM $\mathcal{M}_{\theta}$ with parameters $\theta \in \mathbb{R}^d$. The fine-tuning is performed on a private dataset $\mathcal{D}_{\text{fine}} = \{ (\mathcal{I}_i, \mathcal{T}_i) \}_{i=1}^N$, where each pair consists of an image $\mathcal{I}_i$ and a text sequence $\mathcal{T}_i = \{ w_{1}, \dots, w_{i} \}$. The primary objective is to adapt $\mathcal{M}_{\theta}$ to downstream vision-language tasks by learning parameters $\theta_{\text{fine}}$ that exhibit \textbf{high utility}. This utility is typically achieved by minimizing an empirical risk, often the negative log-likelihood loss, over the $\mathcal{D}_{\text{fine}}$.

A crucial and defining requirement for this process is that it must adhere to a \textbf{strict $(\epsilon, \delta)$-Differential Privacy (DP) guarantee} (Definition~\ref{def:dp}) with respect to $\mathcal{D}_{\text{fine}}$. 
This requires the learning algorithm $\mathcal{A}$ to generate $\theta_{\text{fine}}$ from $\mathcal{D}_{\text{fine}}$ and $\theta$ under $(\epsilon, \delta)$-DP guarantees. The core problem can be summarized as finding parameters $\theta_{\text{fine}}$ that balance utility and privacy, as formally stated below:

\begin{tcolorbox}[
    colback=gray!10,  
    colframe=black!95, 
    arc=4pt,          
    boxrule=1pt,    
    title=Problem Formulation, 
    fonttitle=\bfseries,
    breakable,        
    enhanced,         
    attach boxed title to top center={yshift=-0.25mm-\tcboxedtitleheight/2,yshifttext=2mm-\tcboxedtitleheight/2},
    boxed title style={colback=gray!75,colframe=gray!75,sharp corners}
]
\textbf{Objective:} Minimize the empirical risk on the private dataset $\mathcal{D}_{\text{fine}}$:
\begin{equation}
\mathcal{L}(\theta, \mathcal{D}_{\text{fine}}) := \frac{1}{N} \sum_{i=1}^N \left( - \sum_{t=1}^{T_i} \log P_{\mathcal{M}_{\theta}}(w_{i,t} \!\mid\! \mathcal{I}_i, w_{i,1}, \dots, w_{i,t-1}) \right)
\label{eq:mllm_loss_in_box} 
\end{equation}
 The learning algorithm $\mathcal{A}$ producing $\theta_{\text{fine}}$ from $\mathcal{D}_{\text{fine}}$ must be $(\epsilon, \delta)$-Differentially Private:
\begin{equation}
\text{Find } \theta_{\text{fine}} \approx \argmin_{\theta \in \mathbb{R}^d} \mathcal{L}(\theta, \mathcal{D}_{\text{fine}}) \quad \text{s.t.} \quad \mathcal{A}(\mathcal{D}_{\text{fine}}) \text{ is } (\epsilon, \delta)\text{-DP}.
\label{eq:dp_constrained_opt_in_box} 
\end{equation}
\end{tcolorbox}

\vspace{-2mm}
\section{Method}
\label{sec:methodology}
\vspace{-3mm}
We introduce \textbf{Dual-Priv Pruning}, the first framework for differential private (DP) fine-tuning of MLLMs, designed to optimize the privacy-utility trade-off. 
\textbf{Mechanism 1} performs attention-based token pruning and fusion to transform the visual input into a compact representation \( \mathcal{V}' \).  
\textbf{Mechanism 2} applies \((\epsilon, \delta)\)-DP to the trainable parameters \( \theta_{\text{train}} \) using DP-SGD (\cref{def:dpsgd}), enhanced with a \textit{gradient-update pruning} strategy to improve utility. This  provides formal \((\epsilon, \delta)\)-DP guarantees for the entire pipeline.
Further details and motivations are in Appendix~\ref{app:stage1_motivation} and Appendix~\ref{app:stage2_motivation}.

\begin{figure*}[t] 
    \centering
    \includegraphics[width=1\textwidth]{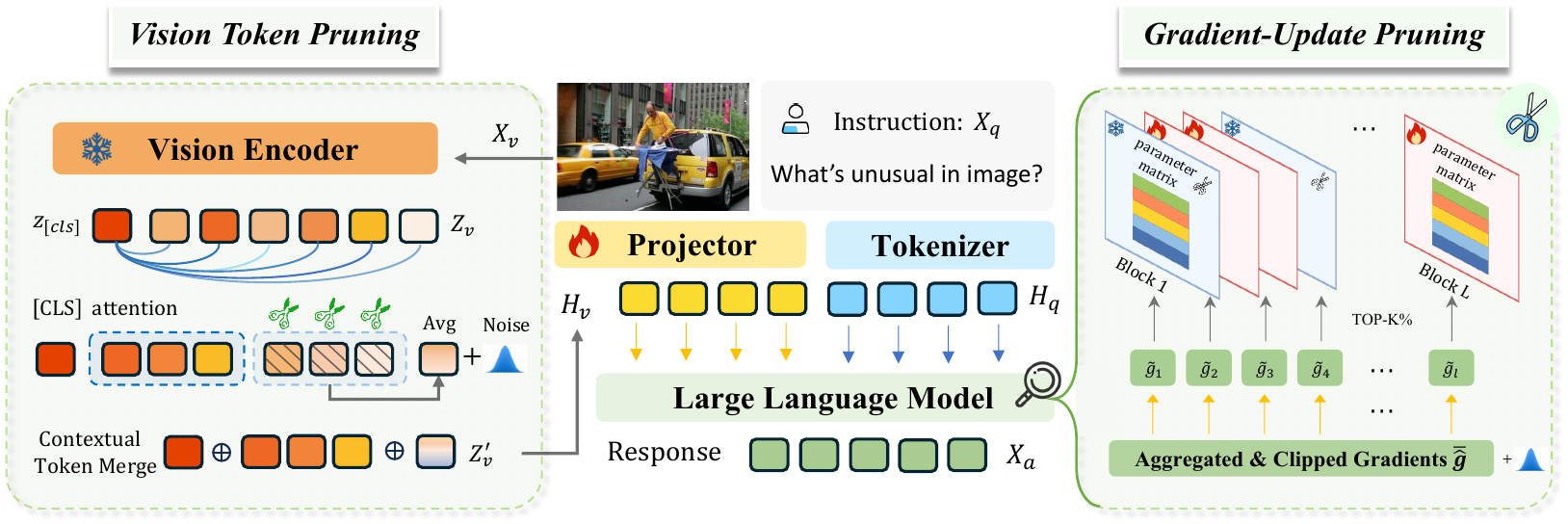} 
    \vspace{-5mm}
    \caption{Overview of our Dual-Priv Pruning. 
    \textbf{(Left)}: Visual Token Pruning and Fusion. Using [CLS] attention, dominant tokens are selected; less important ones are averaged with heuristic noise. 
    \textbf{(Right)}: DP Fine-tuning with gradient pruning. Noise is added to gradients in LLM blocks, and updates are selectively applied based on noisy gradient magnitude. Frozen parameters remain unchanged.}
    \label{fig:pipeline}
    \vspace{-4mm}
\end{figure*}
\subsection{Mechanism 1: Visual Token Pruning and Fusion}\label{stage1}
\vspace{-2mm}
This initial stage reduces the computation cost associated with long visual token sequences before the differential private fine-tuning process begins. It consists of identifying and retaining the most important visual tokens based on attention, followed by merging the remaining tokens and applying noise prior. This stage is not designed to provide the formal DP guarantee.

\noindent \textbf{Dominant Token Selection via CLS Attention.} 
For an input image \(\mathcal{I}_i\), the vision encoder extracts an initial set of \(n\) visual tokens \(\mathcal{V} = \{v_{cls}, v_1, \dots, v_{n-1}\}\), including a class token \(v_{cls}\) and \(n-1\) patch tokens, where \(v_j \in \mathbb{R}^d\). We hypothesize that tokens receiving significant attention from the class token ([CLS]) include the most critical global information.

To identify these dominant tokens, we first compute the multi-head self-attention maps within a selected layer of the vision encoder. The attention map for a single head \(h\) is given by:
\begin{equation}
    S_h = \text{Softmax}\left(\frac{Q_h K_h^\top}{\sqrt{D_h}}\right) \in \mathbb{R}^{n \times n}, \label{eq:single_head_attention_revised}
\end{equation}
where \(Q_h, K_h\) are the query and key matrices, and \(D_h\) is the head dimension. We average these maps across all \(H\) heads to get an aggregated attention map \(S_{\text{avg}} \in \mathbb{R}^{n \times n}\):
\begin{equation}
S_{\text{avg}} = \frac{1}{H} \sum_{h=1}^H S_h. \label{eq:avg_attention_revised}
\end{equation}
The importance score \(s_j\) for each patch token \(v_j\) (\(j \in \{1, \dots, n-1\}\)) is then determined by the attention receives from the [CLS] token in the aggregated map. We select the \(K\) patch tokens with the highest importance scores \(s_j\) as the dominant patch tokens \(\mathcal{V}_d = \{ v_j \mid s_j \text{ is among the top K scores} \}\). The class token \(v_{cls}\) is always retained. The remaining patch tokens form the non-dominant set \(\mathcal{V}_{nd}\).

\noindent \textbf{Contextual Token Fusion and Heuristic Noise.} 
\label{sec:contextual_merging_heuristic}To preserve the visual context features from \(\mathcal{V}_{nd}\) while reducing sequence length, we uniformly randomly select tokens \( v_{\text{center},i} \) from \(\mathcal{V}_{nd}\) as cluster centers and enhance their representation based on cosine similarity with the remaining non-dominant tokens. Subsequently, Gaussian noise scaled by \(\sigma_{\text{fuse}}^2\) is heuristically applied to the enhanced \( v_{\text{center}} \), producing the fused contextual tokens \(c\), as defined in the following formula:
\begin{equation}
\label{fus_final}
\mathbf{c} = \begin{bmatrix}
v_{\text{center},1} + \frac{1}{|\mathcal{C}_1|} \sum_{v_j \in \mathcal{C}_1} v_j \\
v_{\text{center},2} + \frac{1}{|\mathcal{C}_2|} \sum_{v_j \in \mathcal{C}_2} v_j \\
\vdots \\
v_{\text{center},k} + \frac{1}{|\mathcal{C}_k|} \sum_{v_j \in \mathcal{C}_k} v_j
\end{bmatrix} + \mathcal{N}\left(0, \sigma_{\text{fuse}}^2 I_{k d}\right),
\end{equation}
$\text{where }\mathcal{C}_i \text{ is the set of non-dominant tokens assigned to the } i\text{-th cluster based on similarity:}$
\begin{align}
&
\quad \mathcal{C}_i = \left\{ v_j \in \mathcal{V}_{nd} \mid i = \arg\max_{l} \text{sim}(v_j, v_{\text{center},l}) \right\}, \quad i = 1, 2, \dots, k.
\end{align}
The noise adding process serves as a form of regularization or stochasticity injection; A key aspect of our design is to maintain consistency with the noise introduced by the DP mechanism in the subsequent stage. Therefore, the variance of this heuristic noise, $\sigma_{\text{fuse}}^2$, is set to be equivalent to the variance of the Gaussian noise added per step in the DP optimization process (Mechanism 2, \cref{stage2:revised}). It does not contribute to the formal $(\epsilon, \delta)$-DP guarantee derived in Mechanism 2. The final set of visual tokens passed to the MLLM for the DP fine-tuning stage is \(\mathcal{V}' = \{v_{cls}\} \cup \mathcal{V}_d \cup \{C\}\), which has a significantly reduced size of \(K+|C|+1\) tokens.
\vspace{-2mm}
\subsection{Mechanism 2: DP Fine-tuning with gradient-update pruning}
\label{stage2:revised}
\vspace{-2mm}
This core mechanism performs the $(\epsilon, \delta)$-differential private fine-tuning of the trainable parameters $\theta_{\text{train}}$ (e.g., LoRA matrices~\cite{hu2022lora}), leveraging the pruned visual inputs $(\mathcal{V}', \mathcal{T})$ from Mechanism 2. Our approach builds upon DP-SGD (\cref{def:dpsgd}) but introduces a \textbf{post-noise adaptive update} mechanism designed to enhance utility without compromising the privacy guarantee.

The process within each training iteration $t$ begins with standard DP-SGD procedures. For a minibatch $\xi_t$ of size $m$, we first compute per-sample gradients $g_i = \nabla_{\theta_{\text{train}}} \mathcal{L}(\theta_{t-1}; (\mathcal{V}'_i, \mathcal{T}_i))$. To bound the influence of individual samples, we clip the $L_2$ norm of each gradient using a threshold $C$: $\hat{g}_i = g_i / \max(1, \|g_i\|_2 / C)$. These clipped gradients are then averaged across the minibatch to produce $\bar{\hat{g}} = \frac{1}{m} \sum_{i \in \xi_t} \hat{g}_i$.
The crucial step for ensuring differential privacy follows. Gaussian noise is added \textit{unconditionally} to the entire aggregated gradient vector:
\begin{equation}
\tilde{g} = \bar{\hat{g}} + \mathcal{N}\left(0, \frac{\sigma^2 C^2}{m^2} I_{d_{\text{train}}}\right).
\label{eq:unconditional_noise_main}
\end{equation}
Here, $d_{\text{train}}$ is the dimensionality of $\theta_{\text{train}}$, and the noise multiplier $\sigma$ is determined by the overall privacy budget $(\epsilon, \delta)$, number of steps $T$, and sampling rate $q$ via privacy accounting (\cref{fact:composition_accounting}). At this point, the noisy gradient $\tilde{g}$ is an $(\epsilon_t, \delta_t)$-differentially private quantity for the current step.
Our mechanism diverges from standard DP-SGD hereafter. Instead of directly using $\tilde{g}$ for the update, we first analyze its structure and magnitude. We partition $\tilde{g}$ into components $\tilde{g}_j$ corresponding to logical parameter blocks within $\theta_{\text{train}}$ and compute the $L_2$ norm $N_j = \|\tilde{g}_j\|_2$ for each block.

Based on these norms, we generate a binary mask $M$, structured identically to $\theta_{\text{train}}$, to selectively prune the parameter update. A block $j$ is chosen for update ($M_j$ remains 1): \textbf{only if its noisy gradient norm $N_j$ is among the top K\% of norms across all blocks}, otherwise $M_j$ remains 0. 
\begin{equation}
M_j = \mathbb{I}(N_j \in \text{Top-K\%}(\{N_1, N_2, \dots, N_J\})),
\label{eq:mask_generation_main_topk} 
\end{equation}
where $\mathbb{I}(\cdot)$ is the indicator function, $J$ is the total number of parameter blocks, and $\text{Top-K\%}(\cdot)$ denotes the set of the $K$\% largest norm values. The percentage for K\% is a hyperparameter.

Finally, the model parameters are updated using the noisy gradient $\tilde{g}$, but applied selectively through the generated mask $M$ via element-wise multiplication (Hadamard product $\odot$):
\begin{equation}
\theta_t = \theta_{t-1} - \eta_t \cdot (M \odot \tilde{g}).
\label{eq:masked_update_main}
\end{equation}
This ensures that parameter updates are only applied to blocks where the noisy gradient signal was deemed sufficiently strong or reliable according to the gating criterion. The full step-by-step procedure is formally detailed in \cref{app:algo_details}. 


\vspace{-2mm}
\subsection{Overall Privacy Guarantee}
\label{sec:overall_privacy}
\vspace{-2mm}
The $(\epsilon, \delta)$-DP guarantee of the Dual-Priv Pruning method is entirely derived from Mechanism 2 (\cref{stage2:revised}). Mechanism 1 (\cref{stage1}) involves data preprocessing \textit{before} the DP mechanism is applied and does not consume privacy budget. The adaptive update mechanism within Mechanism 2, constitutes post-processing on the private intermediate result $\tilde{g}$ and thus does not affect the formal $(\epsilon, \delta)$-DP guarantee (\cref{app:post_processing}).

\vspace{-2mm}
\section{Experiments}
\label{sec:experiments}
\vspace{-3mm}
We conduct a comprehensive experimental evaluation of our proposed \textbf{Dual-Priv Pruning} method. Our experiments are designed to validate four core advantages of Dual-Priv Pruning: \textbf{(1)} Preserve utility, especially under strict privacy budgets ($\epsilon \le 3$), compared to baseline methods; \textbf{(2)} Significant improvements in computation cost, highlighted by an approximate 14.34\% reduction in peak GPU memory usage; and \textbf{(3)} Validated effectiveness on challenging, visual tasks, encompassing high-resolution real-world scenes and medical images, demonstrating the method's practical applicability in complex, privacy-sensitive domains. \textbf{(4)}  Empirically shown to be effective against privacy attacks like Membership Inference Attacks (MIA).

\subsection{Experimental Setup}
\label{subsec:setup}
\vspace{-2mm}
\textbf{Datasets.} 
 We evaluate performance by fine-tuning on the training sets and evaluating on the test sets of several vision-language benchmarks. These include standard datasets such as ScienceQA~\cite{lu2022learn} (Scientific VQA), TextVQA~\cite{singh2019towards} (VQA over text in images), and GQA~\cite{hudson2019gqa} (Compositional VQA). To specifically assess scalability and robustness on complex inputs, we utilize MME-RealWorld~\cite{zhang2024mme}, an MLLM benchmark designed for high-difficulty tasks involving high-resolution real-world images. Additionally, we incorporate two medical visual question answering dataset, PathVQA~\cite{he2020pathvqa}and VQA-RAD~\cite{lau2018dataset}, to further test generalization on specialized, challenging domains. 
 
\textbf{Model \& Training Strategy.} We utilize LLAVA-7B~\cite{liu2023visual} as our base MLLM. Specifically, for tasks in the medical domain (PathVQA, VQA-RAD, and MIA on ROCOV2), we employ Med-LLaVA\cite{li2023llava}, a LLaVA variant adapted for medical vision-language understanding. To isolate the impact of DP fine-tuning methods, we do not perform additional instruction tuning stages beyond the initial pre-training of LLAVA. Parameter-efficient fine-tuning is achieved using LoRA~\cite{hu2022lora} (rank $r=128$, scaling $\alpha=256$) with batch size 12. All models are trained on the train set using the Adam optimizer~\cite{kingma2014adam} with a learning rate of 2e-4 for 1 epoch. We use 4 A100
40G GPUs for training.

\textbf{DP Implementation.} We guarantee $(\epsilon, \delta)$-DP via the Gaussian Mechanism Privacy loss is tracked using Rényi Differential Privacy(RDP)~\cite{DBLP:conf/csfw/Mironov17}. We set $\delta$ close to the inverse dataset size ($1/N$) and evaluate across strict to mild privacy budgets: $\epsilon \in \{1, 3, 8\}$. Per-sample gradients are clipped at a maximum $L_2$ norm of $C=1.0$.

\textbf{Baselines.} Our Dual-Priv Pruning method is compared against:
 DP-SGD~\cite{abadi2016deep}: The standard baseline for DP fine-tuning, applying  Gaussian noise to the averaged clipped gradients of all trainable parameters. DPZO~\cite{tang2024private}: A representative zeroth-order DP optimization method, included to assess alternatives that avoid direct gradient computation. Detailed for baselines are in Appendix~\ref{app:baseline_details}. 

\textbf{Dual-Priv Pruning Configuration.} 
Mechanism 1 (\cref{stage1}) retains $K=191$ attention-selected visual tokens plus [CLS] and 30 fused token (40\% of total).
Mechanism 2 (\cref{stage2:revised}) employs gradient-update pruning by selecting parameter blocks for update if their noisy gradient norms are among the \textbf{top 80\%} of all block norms (\cref{eq:mask_generation_main_topk}).

\vspace{-2mm}
\subsection{Performance on Standard Benchmarks}
\vspace{-2mm}

\label{subsec:standard_results}
\begin{table*}[t!]
    \centering
    \caption{
    Comparison of different methods on standard benchmarks (BS = 12). For reference, non-private performance ($\epsilon=\infty$) are included. Metrics reported are Accuracy (Acc) and Image-based Accuracy (IMG). The best results for each $\epsilon$ setting are shown in \textbf{bold}.
    }
    \vspace{-2mm}
    \label{tab:1}
    \footnotesize
    \renewcommand{\arraystretch}{1.3} 
    \setlength{\tabcolsep}{3pt} 
    \sisetup{round-mode=places, round-precision=4, table-format=1.4} 
    
    \scalebox{0.9}{
    \begin{tabular}{@{} >{\centering\arraybackslash}p{0.6cm} | cccc | cccc | cccc @{}}
        \toprule
        \multirow{3.5}{*}{\makebox[0.6cm][c]{$\epsilon$}} & \multicolumn{4}{c|}{DZPO} & \multicolumn{4}{c|}{DP-SGD} & \multicolumn{4}{c@{}}{Dual-Priv(ours)} \\
        \cmidrule(lr){2-5} \cmidrule(lr){6-9} \cmidrule(l){10-13}
        & \multicolumn{2}{c}{ScienceQA} & {TextVQA} &{GQA} & \multicolumn{2}{c}{ScienceQA} & {TextVQA} &{GQA} & \multicolumn{2}{c}{ScienceQA} & {TextVQA} &{GQA} \\
        \cmidrule(lr){2-3} \cmidrule(lr){4-4}\cmidrule(lr){5-5} \cmidrule(lr){6-7} \cmidrule(lr){8-8} \cmidrule(lr){9-9} \cmidrule(lr){10-11} \cmidrule(lr){12-12} \cmidrule(l){13-13}
        & {Acc(\%)} & {IMG} & {Acc(\%)} & {Acc(\%)} & {Acc(\%)} & {IMG} & {Acc(\%)} & {Acc(\%)} & {Acc(\%)} & {IMG} & {Acc(\%)} & {Acc(\%)} \\
        \midrule
        1 & 23.30 & 21.50 & 1.13 & 0.00 & 81.54 & 72.51 & 34.52 & 38.61 & \cellcolor{gray!12}\textbf{84.20} & \cellcolor{gray!12}\textbf{78.43} & \cellcolor{gray!12}\textbf{34.74} & \cellcolor{gray!12}\textbf{39.06} \\
        3 & 21.50 & 19.90 & 2.82 & 0.00 & 78.80 & 70.59 & \cellcolor{gray!12}\textbf{35.64} & 39.11 & \cellcolor{gray!12}\textbf{82.80} & \cellcolor{gray!12}\textbf{75.98} & 35.17 & \cellcolor{gray!12}\textbf{39.65} \\
        8 & 21.50 & 19.90 & 1.31 & 0.00 & 82.52 & 74.00  & 35.60 & 39.16 &\cellcolor{gray!12}\textbf{85.10} & \cellcolor{gray!12}\textbf{76.47} & \cellcolor{gray!12}\textbf{35.71} &\cellcolor{gray!12}\textbf{39.78} \\
        $\infty$ &22.16 & 0.98 & 0.95 & 0.00 & 81.10 & 73.53 & 34.89 & 38.92 & \cellcolor{gray!12}\textbf{84.60} & \cellcolor{gray!12}\textbf{79.41} & \cellcolor{gray!12}\textbf{35.53} & \cellcolor{gray!12}\textbf{39.06} \\
        \bottomrule
    \end{tabular}}
    \vspace{-2mm}
\end{table*}

 The results presented in Table~\ref{tab:1} demonstrate the efficacy of Dual-Priv Pruning. Our method consistently achieves performance that is often superior to DP-SGD, especially under stricter privacy constraints ($\epsilon \!\in\! \{1, 3\}$). DPZO consistently underperforms across all settings, yielding significantly lower accuracy (e.g., only 23.30 on ScienceQA at $\epsilon\!=\!1$, and 0.00 on GQA). This poor performance is largely attributed to the significant convergence difficulties encountered when applying zeroth-order optimization directly to complex MLLM training under DP constraints. On ScienceQA, our approach excels. At the budget of $\epsilon\!=\!3$, Dual-Priv Pruning achieves \textbf{82.80} and a notable \textbf{75.98} IMG, much outperforming DP-SGD (78.80/70.59 respectively). This significant gain, particularly in the visual-dependent IMG metric, underscores our method's strength in preserving vital visual information despite DP noise. This performance likely benefits from the synergistic effect of visual token pruning and fusion and selective gradient-update. Even at the strictest budget of $\epsilon\!=\!1$, our method maintains a clear advantage on ScienceQA (84.20 vs 81.54; 78.43 vs 72.51). And our method achieves the best non-private performance ($\epsilon=\infty$).On TextVQA and GQA, our method generally performs slightly better than DP-SGD across various privacy levels, confirming its broad applicability. For instance, on GQA, we achieve the best DP performance across all tested $\epsilon$ values. On TextVQA, performance is highly competitive. While DP-SGD leads slightly at $\epsilon\!=\!3$ (35.64 vs 35.17), our method achieves the highest accuracy at $\epsilon\!=\!1$ and $\epsilon\!=\!8$ and also obtains the best non-private result. These results suggest that Dual-Priv Pruning offers a more robust privacy-utility trade-off than standard DP techniques for MLLMs. Its strengths are particularly notable under tight privacy budgets.

\vspace{-2mm}
\subsection{Performance on Medical Visual Tasks}
\label{subsec:medical_results_full} 
\vspace{-2mm}
To further assess applicability in privacy-sensitive domains, we evaluated performance on PathVQA~\cite{he2020pathvqa} (pathology) and VQA-RAD~\cite{lau2018dataset} (radiology). Table~\ref{tab:medical_combined_dp_adjusted_eps_col_ours_right} presents a detailed comparison of performance under different privacy budget.
\begin{table*}[tbp]
    \centering
    \caption{
    Comparison on PathVQA and VQA-RAD under different DP budgets (ours on the right).
}
\label{tab:medical_combined_dp_adjusted_eps_col_ours_right}
    \vspace{-2mm}
    \footnotesize
    \renewcommand{\arraystretch}{1.3} 
    \setlength{\tabcolsep}{3pt} 
    \sisetup{round-mode=places, round-precision=4, table-format=1.4} 
    
    \scalebox{0.8}{%
    \begin{tabular}{@{} >{\centering\arraybackslash}p{0.6cm} | SSS S[table-format=2.1,round-precision=1] | SSS S[table-format=2.1,round-precision=1] | SSS S[table-format=2.1,round-precision=1]@{}}
        \toprule
        \multirow{3.5}{*}{\makebox[0.6cm][c]{$\epsilon$}} & \multicolumn{4}{c|}{DPZO} & \multicolumn{4}{c|}{DP-SGD} & \multicolumn{4}{c@{}}{Ours (Dual-Priv)} \\ 
        \cmidrule(lr){2-5} \cmidrule(lr){6-9} \cmidrule(l){10-13} 
        & \multicolumn{3}{c}{PathVQA} & {VQA-RAD} & \multicolumn{3}{c}{PathVQA} & {VQA-RAD} & \multicolumn{3}{c}{PathVQA} & {VQA-RAD} \\ 
        \cmidrule(lr){2-4} \cmidrule(lr){5-5} \cmidrule(lr){6-8} \cmidrule(lr){9-9} \cmidrule(lr){10-12} \cmidrule(l){13-13} 
        & {BLUE} & {EXT} & {F1} & {Acc(\%)} & {BLUE} & {EXT} & {F1} & {Acc(\%)} & {BLUE} & {EXT} & {F1} & {Acc(\%)} \\ 
        \midrule
        1 & 0.6534 & 0.0301 & 0.0592 & 0.0 & 0.7222 & 0.3732 & 0.3675 & 47.3 & \cellcolor{gray!12}\bfseries 0.7385 & \cellcolor{gray!12}\bfseries 0.3840 & \cellcolor{gray!12}\bfseries 0.3792 & \cellcolor{gray!12}\bfseries 48.6 \\
        3 & 0.6534 & 0.0301 & 0.0592 & 0.0 & 0.7257 & 0.3712 & 0.3653 & 48.1 & \cellcolor{gray!12}\bfseries 0.7263 & \cellcolor{gray!12}\bfseries 0.3738 & \cellcolor{gray!12}\bfseries 0.3701 & \cellcolor{gray!12}\bfseries 48.8 \\
        8 & 0.6534 & 0.0301 & 0.0592 & 0.0 & 0.7140 & 0.3683 & 0.3635 & 46.8 & \cellcolor{gray!12}\bfseries 0.7195 & \cellcolor{gray!12}\bfseries 0.3763 & \cellcolor{gray!12}\bfseries 0.3713 & \cellcolor{gray!12}\bfseries 49.0 \\
        \bottomrule
    \end{tabular}%
    } 
    \vspace{-4mm}
\end{table*}
Our method, consistently outperformed DP-SGD across all the metrics, particularly under stricter privacy budgets. For $\epsilon=1$:
on PathVQA, our approach achieved scores of \num{0.7385}(BLUE), \num{0.384} (EXT), and \num{0.3792} (F1), compared to DP-SGD's \num{0.7222}, \num{0.3732}, and \num{0.3675}, respectively.
On VQA-RAD, our method achieved an accuracy of \num{48.6}\%, surpassing DP-SGD (\num{47.3}\%).
The DPZO baseline performed poorly on both medical datasets.
These consistent gains underscore the potential of Dual-Priv Pruning for tuning MLLMs on sensitive medical data while effectively balancing privacy and utility. 

\vspace{-2mm}
\subsection{Performance on High Resolution Visual Tasks} 
\label{subsec:mme_realworld_results} 
\vspace{-2mm}
\begin{wrapfigure}[8]{r}{0.5\columnwidth} 
\vspace{-3mm}
    \setlength{\intextsep}{1pt}           
    \centering
    \sisetup{
        round-mode=places,
        round-precision=2,
        table-format=2.2,                 
        detect-weight                     
    }
    \captionof{table}{Accuracy (\%) on the MME-RealWorld Benchmark (Lite version evaluation, BS=12). 
    }
\label{tab:mme_realworld_wrap}
\vspace{-2mm}
    \setlength{\tabcolsep}{3pt}
    \scalebox{0.9}{\begin{tabular}{@{}l S S S S@{}}
        \toprule
        Method                 & {$\epsilon=1$} & {$\epsilon=3$} & {$\epsilon=8$} & {$\epsilon=\infty$} \\
        \midrule
        DPZO                   & 0.89           & 19.80          & 6.33           & 22.67 \\
        DP-SGD                 & 35.44          & 44.03          & 42.17          & 44.50 \\
        Ours (Dual-Priv)       & \bfseries \cellcolor{gray!12}43.98 & \bfseries \cellcolor{gray!12}45.34 & \bfseries \cellcolor{gray!12}44.40 &  42.16 \\ 
        \bottomrule
    \end{tabular}}
\end{wrapfigure}
To evaluate performance on tasks demanding fine-grained perception and complex reasoning crucial for real-world applicability, we utilize the MME-RealWorld benchmark~\cite{zhang2024mme}. Applying differential privacy in such scenarios is particularly challenging, as the noise required for privacy can significantly impair the model's ability to discern visual details and perform nuanced reasoning.
Models are DP-finetuned on the main MME-RealWorld training dataset and subsequently evaluated on the held-out MME-RealWorld \textbf{lite version}. The results, presented in \cref{tab:mme_realworld_wrap}, demonstrate a substantial advantage for Dual-Priv Pruning over both DP-SGD and DPZO across all tested privacy budgets ($\epsilon \in \{1, 3, 8\}$). 
Notably, under the strict $\epsilon=1$ setting, our method achieves \textbf{43.98 }accuracy, significantly surpassing DP-SGD (\textbf{35.44}).
The significant performance gain on this challenging benchmark underscores the effectiveness of Dual-Priv Pruning. Our method, appears better equipped to preserve the crucial reasoning capabilities, even under stringent privacy constraints. This suggests Dual-Priv Pruning is a promising approach for deploying privacy-preserving MLLMs in real-world applications demanding high visual fidelity and complex reasoning.

\vspace{-2mm}
\subsection{Computational Efficiency Analysis}
\label{subsec:efficiency_revised}
\begin{figure*}[t] 

    \centering 

    \includegraphics[width=1.0\textwidth]{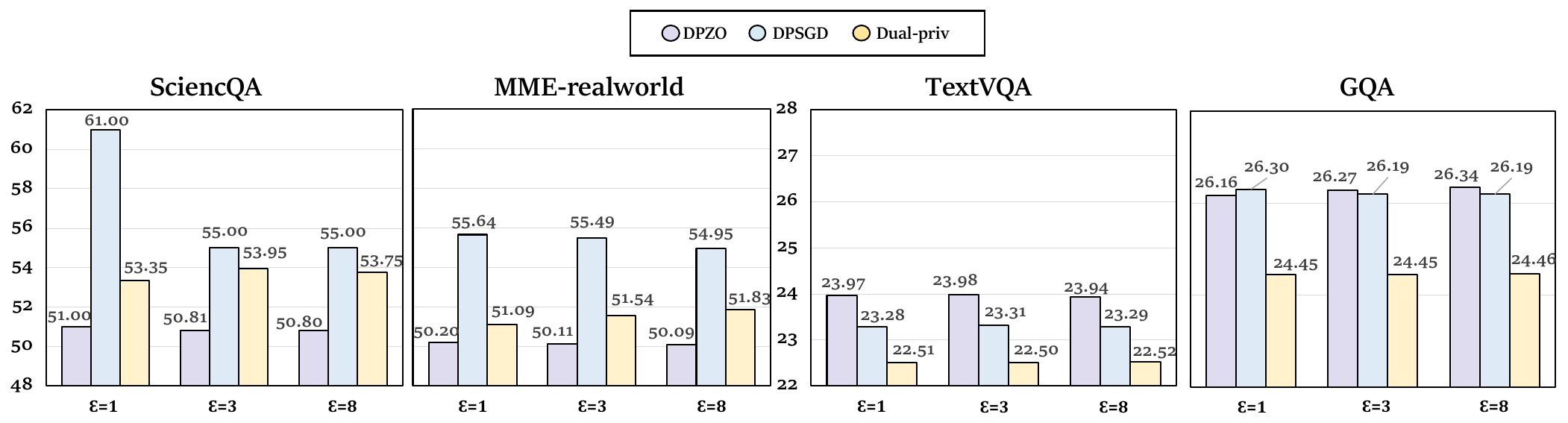} 


    \caption{Average GPU memory consumption (in GB) during fine-tuning for DPZO, DP-SGD, and our Dual-Priv Pruning across four datasets: ScienceQA, MME-RealWorld (evaluated on 4xA100 40G GPUs), and TextVQA, GQA (evaluated on a single H20 96GB GPU). Experiments were conducted with varying privacy budgets ($\epsilon \in \{1, 3, 8\}$). Lower bars indicate greater memory efficiency.}
    \label{fig:main_performance_compariso} 

    \vspace{-5mm} 
\end{figure*}

\vspace{-2mm}
Figure~\ref{fig:main_performance_compariso} illustrates the average GPU memory usage during fine-tuning for our method compared to the baselines. Across the evaluated datasets, scienceqa on 4 A100s, Dual-Priv Pruning achieves an average \textbf{reduction in average GPU memory usage of approximately 14.34\%}. Although DPZO slightly reduces 1.74\% GPU memory compared with our approach. ( It costs 16.7\% more time per training step and causes a 56.3\% performance loss ). But during tested in H20, our method achieve the lowest consumption of GPU memory. This highlights Dual-Priv Pruning's strength in achieving a favorable balance between model performance and robust computational efficiency, thereby making DP fine-tuning for MLLMs more practical. Dual-Priv enables the DP fine-tuning of MLLMs with more constrained resources.

\vspace{-2mm}
\subsection{Ablation Study}
\label{subsec:ablation_revised_wrap} 
\vspace{-2mm}


\begin{wrapfigure}[7]{r}{0.37\columnwidth} 
    \vspace{-5mm}
    \setlength{\intextsep}{0pt} 

    \sisetup{detect-weight, mode=text} 
    \captionof{table}{Ablation on ScienceQA.}
    \label{tab:ablation_wrap} 
    \vspace{-2mm}
    \centering
    \small 
    \setlength{\tabcolsep}{1pt} 
    \renewcommand{\arraystretch}{0.8} 
    \begin{tabular}{@{}l S[table-format=2.2] S[table-format=2.2]@{}} 
        \toprule
        Configuration & {ACC} & {IMG} \\ 
        \midrule
        Full Method          & \bfseries 84.20 & \bfseries 78.43 \\
        \midrule
        w/o Fusion Noise     &83.50  &76.47  \\
        Mechanism 2 Only         &83.00  &76.96  \\
        Mechanism 1 + Uniform DP & 82.80 & 74.51 \\
        \bottomrule
    \end{tabular}
\end{wrapfigure} 
Ablation results on ScienceQA ($\epsilon=1$) are presented in Table~\ref{tab:ablation_wrap}. The "w/o Fusion Noise" setting, which omits the input-level noise, shows performance decrease (83.50/76.47 ACC/IMG) compared to the Full Method (84.20/78.43 ACC/IMG). This suggests that our strategy of preconditioning the input with noise consistent with the DP optimization offers a beneficial, albeit auxiliary, contribution to performance, without adding a tunable hyperparameter for this noise. Omitting Mechanism 1's token pruning entirely (Mechanism 2 Only) lowers accuracy to 83.00/76.96 and eliminates the computational efficiency benefits. Furthermore, replacing Mechanism 2's selective update with uniform DP-SGD noise significantly degrades performance to 82.80/74.51, confirming the effectiveness of our adaptive update strategy.  
These findings demonstrate that both Mechanism 1 and Mechanism 2 are crucial components to the overall performance of dual-private pruning.
\subsection{Impacts of Pruning Ratios}
\label{subsec:pruning_ratio} 
\vspace{-2mm}

\begin{wrapfigure}{r}{0.5\textwidth} 
    \vspace{-14mm} 
    \centering
    \includegraphics[width=0.48\textwidth]{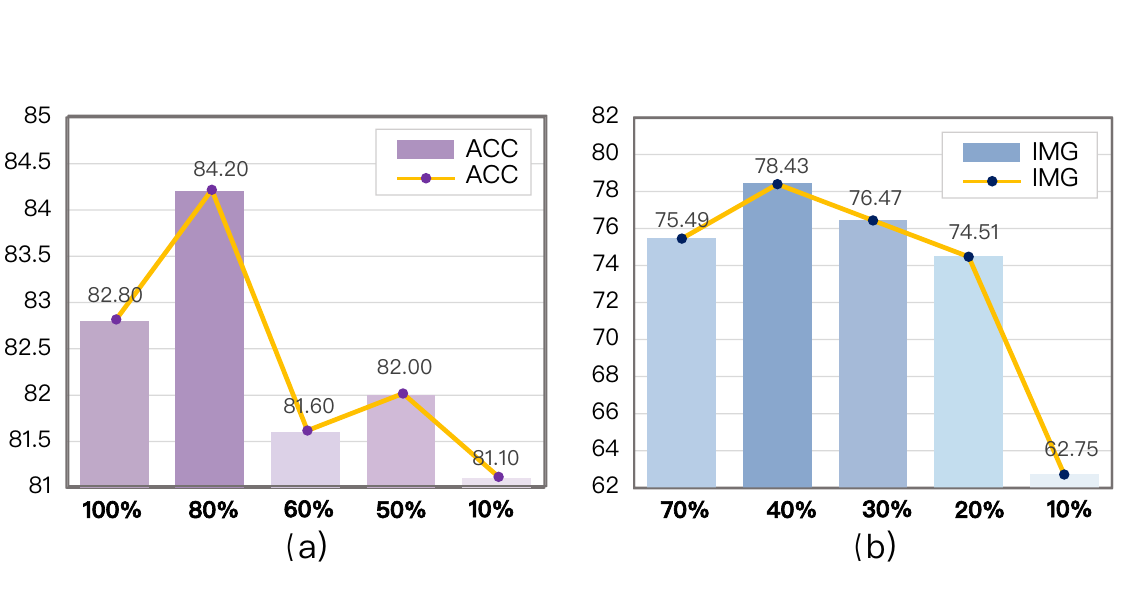} 
    \vspace{-3mm} 
    \caption{Pruning ratios impacts on ScienceQA  ($\epsilon=1$). (a) percentage of top K\% gradient blocks updated (Mechanism 2). (b) percentage of visual tokens retained (Mechanism 1).} 
    \label{fig:pruning_ratio_influence} 
    \vspace{-4mm} 
\end{wrapfigure}

We examine the impact of different pruning ratios within the Dual-Priv Pruning framework on the ScienceQA dataset ($\epsilon=1$).
\cref{fig:pruning_ratio_influence} (a) displays the relationship between the \textbf{gradient-update pruning ratio} and overall accuracy (ACC). The ACC peaks at \textbf{84.20} when the top 80\% of blocks are updated. Updating all blocks yields a lower ACC of 82.80. Reducing the update ratio further (60\%, 50\%, 10\%) leads to ACC values of 81.60, 82.00, and 81.10.\cref{fig:pruning_ratio_influence} (b) illustrates how the \textbf{visual token retention ratio} affects image-based accuracy (IMG). The IMG shows a peak of \textbf{78.43\%} when retaining 40\% of the visual tokens. Retaining more tokens (e.g., 70\%) results in a lower IMG of 75.49, while retaining fewer tokens progressively reduces performance (76.47 at 30\%, 74.51 at 20\%), with a sharp drop to 62.75 when only 10\% of tokens are kept. 
These experiments highlight that both pruning mechanisms involve a delicate trade-off. Optimal performance requires retaining sufficient signal (visual features or gradient updates) while pruning elements that might be redundant or overly affected by DP noise. 

\subsection{Performance Under Membership Inference Attacks}
\label{subsec:mia_results}
\vspace{-2mm}
\begin{wrapfigure}{r}{0.4\textwidth} 
    \vspace{-5mm} 
    \centering
    \includegraphics[width=0.4\textwidth]{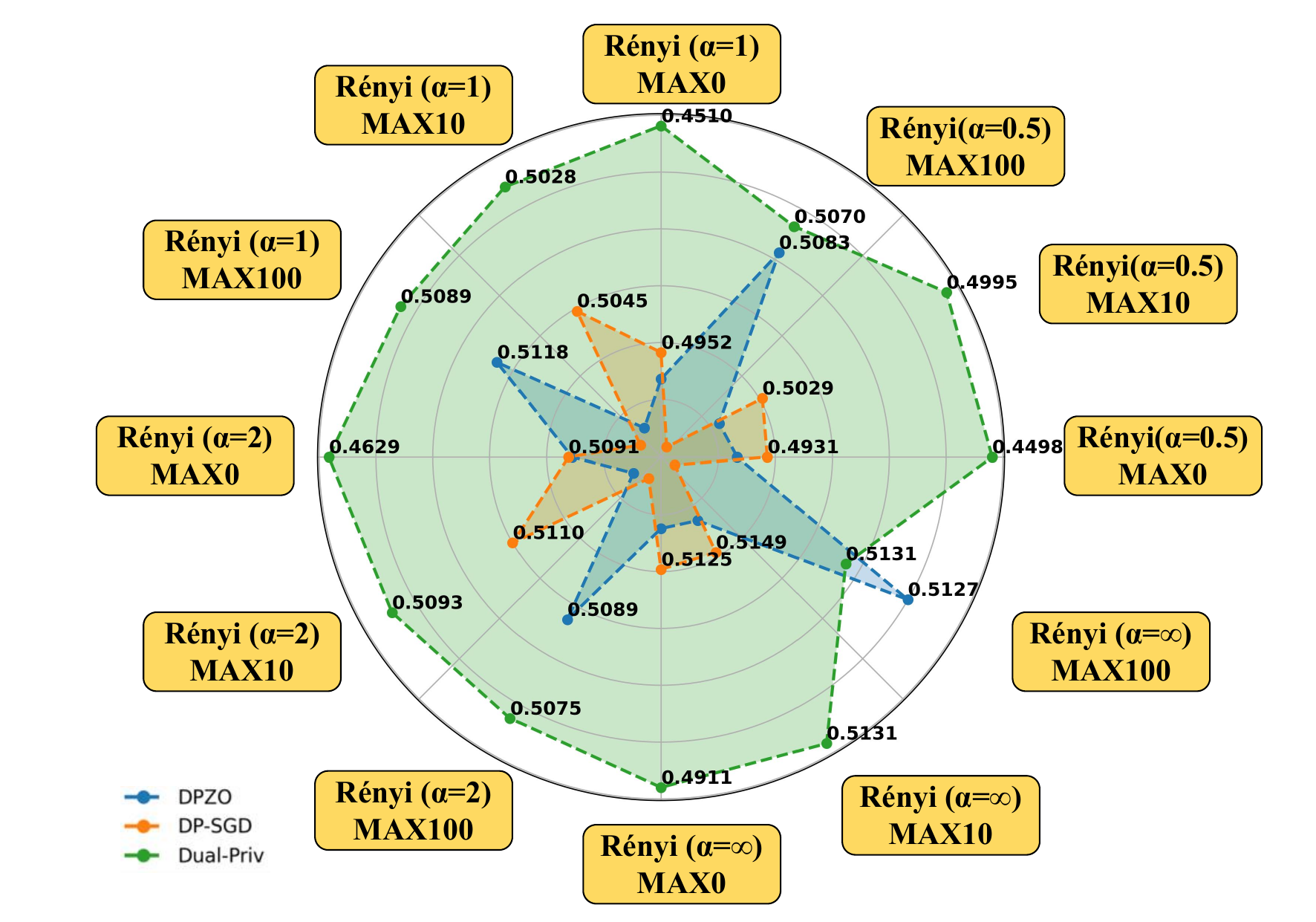} 
    \vspace{-5mm}
    \caption{Radar chart of AUC under varying Rényi entropy orders and top entropy percentages. Metrics use strict privacy budget ($\epsilon\!=\!1$). Distribution places smaller values near edges.}
    \label{MIA_radar}
    \vspace{-10pt} 
\end{wrapfigure}
To further test the privacy protection capability of our approach, we validate the performance through membership inference attack\cite{shokri2017membership}.
The latest MIA design for MLLM \cite{li2024membership} was adopted as the evaluation pipeline.
Models were DP-finetuned on privacy sensitive medical image caption dataset \textbf{ROCOV2}\cite{ruckert2024rocov2} with the batchsize of 12 following the standard setup (\cref{subsec:setup}).
Extensive experiments demonstrate that our work outperform both DPZO and DPSGD across metrics include AUC and ACC, especially in protecting visual information as it benefit from adding heuristic fusion noise in Mechanism 1. As it is shown in Figure~\ref{MIA_radar}, the AUC obtained by attacking our model is the lowest under almost every order of Rényi entropy and percentage of top entropies selected, which means that the possibility of distinguishing the member from database is the lowest for Dual-Priv Pruning under the same attack pipeline compared with other methods.
Further data suggests that our method has a strong ability to protect MLLM from assigning a higher "membership score" to a randomly chosen member than to a randomly chosen non-member, which makes Membership inference attack nearly approaches random guessing. Additional details are in Appendix~\ref{app:mia_results}.

\vspace{-2mm}
\section{Conclusion}
\label{subsec:discussion_revised}
In this work, we introduced \textbf{Dual-Priv Pruning}, the first framework for efficient differential private fine-tuning of Multimodal Large Language Models. Our approach combines \textit{visual token pruning} with an input noise strategy aligned with DP noise intensity, and a \textit{gradient-update pruning} mechanism. Extensive experiments demonstrate that Dual-Priv Pruning achieves a compelling privacy-utility trade-off, significantly reducing computational overhead while maintaining competitive performance, especially under stringent privacy budgets. This work represents a crucial first step towards practical privacy-preserving MLLM deployment.
\vspace{-2mm}

\newpage
\bibliographystyle{plainnat}
\bibliography{nip2025}

\clearpage
\appendix
\section{Key Differential Privacy Facts} 
\label{app:dp_facts}
The following facts elucidate key DP properties essential for MLLM fine-tuning:

\begin{fact}[Sensitivity and the Gaussian Mechanism~\cite{dwork2014algorithmic}] 
\label{fact:gaussian_mechanism}
To protect the output of a function \(f\) using noise, we first need to leverage its $L_2$ sensitivity $\Delta f$. This measures the maximum possible change $\|f(\mathcal{D}) - f(\mathcal{D}')\|_2$ when the input dataset changes by one record. If $\Delta f$ is known (or bounded), the Gaussian Mechanism adds noise $\mathcal{N}(0, \sigma_{GM}^2 I)$ where the standard deviation $\sigma_{GM}$ is related to $\Delta f$ and depends on the desired single-step privacy $(\epsilon, \delta)$, calculated as:
\begin{equation}
\sigma_{GM} \ge \frac{\Delta f \sqrt{2 \ln(1.25 / \delta)}}{\epsilon}. \label{eq:gaussian_std_calc} 
\end{equation}
\end{fact}

\begin{fact}[Privacy Accounting via RDP~\cite{DBLP:conf/csfw/Mironov17}] 
\label{fact:composition_accounting} 
Privacy accounting methods are essential for tracking this privacy loss. Rényi Differential Privacy (RDP)~\cite{DBLP:conf/csfw/Mironov17} is widely used for such accounting~\cite{abadi2016deep}. The RDP accountant's practical role is to, given a target overall privacy budget $(\epsilon, \delta)$, total training steps $T$, and sampling rate $q$, compute the required per-step noise multiplier $\sigma$ (~\ref{def:dpsgd}) and suggest a clipping norm $C$ to meet the $(\epsilon, \delta)$-DP guarantee. Mathematical details are in Appendix~\ref{app:rdp_details}.
\end{fact}
\section{Sensitivity Analysis for DP-SGD under Add-or-Remove Adjacency}
\label{app:sensitivity_derivation_content} 

In DP-SGD (Definition~\ref{def:dpsgd}), we apply the Gaussian Mechanism to the average of per-sample clipped gradients. The choice of adjacency definition for datasets $\mathcal{D}$ and $\mathcal{D}'$ (i.e., how they differ by "one record") impacts the sensitivity calculation. As stated in Definition~\ref{def:dp}, our work considers add-or-remove adjacency, where neighboring datasets differ by the addition or removal of a single image-text pair.

Consider the function $f(\mathcal{D}, \theta) = \sum_{x_i \in \mathcal{D}} \hat{g}_i(\theta)$, which is the sum of clipped gradients $\hat{g}_i$ for all samples $x_i$ in a dataset $\mathcal{D}$. Each per-sample clipped gradient satisfies $\|\hat{g}_i\|_2 \le C$.
If we consider two neighboring datasets $\mathcal{D}$ and $\mathcal{D}'$ where $\mathcal{D}' = \mathcal{D} \cup \{x^*\}$ (i.e., $x^*$ is added), then:
\[ \|f(\mathcal{D}, \theta) - f(\mathcal{D}', \theta)\|_2 = \| \sum_{x_i \in \mathcal{D}} \hat{g}_i(\theta) - (\sum_{x_i \in \mathcal{D}} \hat{g}_i(\theta) + \hat{g}_{x^*}(\theta)) \|_2 = \|-\hat{g}_{x^*}(\theta)\|_2 = \|\hat{g}_{x^*}(\theta)\|_2 \le C. \]
Similarly, if $\mathcal{D}' = \mathcal{D} \setminus \{x^*\}$ (i.e., $x^*$ is removed), the difference is also bounded by $C$.
Thus, the $L_2$ sensitivity of the sum of clipped gradients is $\Delta_2 f = C$.

In DP-SGD, we typically compute gradients over a minibatch $\xi_t$ of size $m$ sampled from the full dataset $\mathcal{D}_{\text{train}}$ (of size $N$) with sampling probability $q = m/N$. The noisy update is applied to the \textit{average} of these clipped gradients: $\bar{g} = \frac{1}{m} \sum_{i \in \xi_t} \hat{g}_i$.
For the per-iteration application of DP-SGD with minibatch sampling, the effective $L_2$ sensitivity of the quantity to which noise is added (the average gradient) is commonly taken as $\frac{C}{m}$ under the add-or-remove model when considering the privacy implications for each individual sample's contribution to this average~\cite{abadi2016deep, dwork2014algorithmic}. More precisely, the clipping ensures that the maximum influence of any single user's data on the sum of gradients in a batch is $C$. When this sum is averaged over $m$ samples, the change due to one user's data (if that user were removed from the entire dataset) can be bounded appropriately, leading to the noise calibration based on $C$.

The noise added in DP-SGD (\cref{eq:dpsgd_update}) is $\mathcal{N}(0, \sigma^2 C^2 I_d / m^2)$. This formulation inherently uses $C$ as the sensitivity for the sum of gradients in the batch if we consider each sample's gradient to be a distinct quantity to be protected, and then this noise is scaled by $1/m$ due to averaging (equivalently, the sensitivity of the average is $C/m$).
The critical point is that clipping each per-sample gradient to $C$ bounds its maximum possible contribution. The privacy analysis with subsampling (accounted for by the RDP accountant) then correctly tracks the privacy loss given this per-sample bound $C$.

Therefore, the clipping norm $C$ directly bounds the $L_2$ norm of each individual's contribution before aggregation and noise addition. The Gaussian Mechanism (Fact~\ref{fact:gaussian_mechanism}) is then applied using this understanding, where the effective sensitivity for the noisy average gradient computation in DP-SGD is appropriately scaled by $C$.
\section{Rényi Differential Privacy (RDP) Accounting}
\label{app:rdp_details}
RDP~\cite{DBLP:conf/csfw/Mironov17} provides a way to track privacy loss using Rényi divergence of order $\alpha$, denoted $D_\alpha(P||Q)$. An algorithm $\mathcal{A}$ is $(\alpha, \rho)$-RDP if for all neighboring datasets $\mathcal{D}, \mathcal{D}'$, $D_\alpha(\mathcal{A}(\mathcal{D}) || \mathcal{A}(\mathcal{D}')) \le \rho$. 
Key properties include:
\begin{itemize}
    \item \textbf{Composition:} If $\mathcal{A}_1$ is $(\alpha, \rho_1)$-RDP and $\mathcal{A}_2$ is $(\alpha, \rho_2)$-RDP, their composition $\mathcal{A}_2 \circ \mathcal{A}_1$ is $(\alpha, \rho_1 + \rho_2)$-RDP. This simplifies tracking loss over multiple steps.
    \item \textbf{Gaussian Mechanism RDP:} Adding $\mathcal{N}(0, \sigma_{GM}^2 I)$ noise to a function with $L_2$ sensitivity $\Delta f$ satisfies $(\alpha, \frac{\alpha (\Delta f)^2}{2 \sigma_{GM}^2})$-RDP for any $\alpha > 1$.
    \item \textbf{Subsampling Amplification:} Sampling a minibatch with rate $q$ before applying a DP mechanism amplifies privacy. RDP provides tight bounds for this, especially for Poisson sampling~\cite{abadi2016deep} and uniform sampling without replacement.
    \item \textbf{Conversion to $(\epsilon, \delta)$-DP:} If an algorithm is $(\alpha, \rho)$-RDP for all $\alpha$ in some range, it satisfies $(\epsilon, \delta)$-DP where $\epsilon = \rho + \frac{\log(1/\delta)}{\alpha - 1}$. We typically optimize over $\alpha$ to find the tightest $\epsilon$ for a given $\delta$.
\end{itemize}
The privacy accountant takes $T, q$, the per-step RDP parameters (derived from the Gaussian mechanism using $C$ and $\sigma$), applies composition and subsampling rules to get the total RDP parameters $(\alpha, \rho_{total})$, and converts this to the final $(\epsilon, \delta)$. It can also work backwards: given target $(\epsilon, \delta)$, $T, q$, find the required $\sigma$.

\section{Post-Processing Property of Differential Privacy}
\label{app:post_processing}

The post-processing property is a fundamental and powerful feature of differential privacy~\cite{dwork2014algorithmic}. It states that applying any arbitrary data-independent computation to the output of a differentially private algorithm does not compromise its privacy guarantee.

\textbf{Formal Statement:} Let $\mathcal{A}: \mathcal{D}^n \to \mathcal{R}$ be an $(\epsilon, \delta)$-differentially private algorithm, where $\mathcal{D}^n$ is the space of possible datasets and $\mathcal{R}$ is the output range. Let $f: \mathcal{R} \to \mathcal{R}'$ be any arbitrary randomized or deterministic function whose computation does not depend on the original private dataset $\mathcal{D}$ (it only takes the output of $\mathcal{A}$ as input). Then the composite algorithm $f \circ \mathcal{A}$ (which first runs $\mathcal{A}$ on the input dataset and then applies $f$ to the result) is also $(\epsilon, \delta)$-differentially private.

\textbf{Intuition:} The privacy guarantee provided by $\mathcal{A}$ ensures that its output $Y = \mathcal{A}(D)$ is already "privacy-safe" – observing $Y$ reveals limited information about any individual in $D$. The function $f$ only gets access to this already protected output $Y$. Since $f$ has no additional access to the original sensitive data $D$, it cannot "undo" the privacy protection or learn anything more about individuals in $D$ than what was already bounded by the $(\epsilon, \delta)$-DP guarantee of $\mathcal{A}$.

\textbf{Proof Sketch:} We want to show that for any neighboring datasets $D, D'$ and any set of outcomes $S' \subseteq \mathcal{R}'$:
\[ \Pr[(f \circ \mathcal{A})(D) \in S'] \leq e^{\epsilon} \cdot \Pr[(f \circ \mathcal{A})(D') \in S'] + \delta \]
Let $Y = \mathcal{A}(D)$ and $Y' = \mathcal{A}(D')$. The event $(f \circ \mathcal{A})(D) \in S'$ means $f(Y) \in S'$. Let $S_f = \{ y \in \mathcal{R} \mid f(y) \in S' \}$ be the set of outputs from $\mathcal{A}$ that $f$ maps into $S'$.
Then, $\Pr[(f \circ \mathcal{A})(D) \in S'] = \Pr[Y \in S_f] = \Pr[\mathcal{A}(D) \in S_f]$.
Similarly, $\Pr[(f \circ \mathcal{A})(D') \in S'] = \Pr[\mathcal{A}(D') \in S_f]$.
Since $\mathcal{A}$ is $(\epsilon, \delta)$-DP and $S_f$ is a valid subset of its output range $\mathcal{R}$, we know from Definition~\ref{eq:dp}:
\[ \Pr[\mathcal{A}(D) \in S_f] \leq e^{\epsilon} \cdot \Pr[\mathcal{A}(D') \in S_f] + \delta \]
Substituting back, we get:
\[ \Pr[(f \circ \mathcal{A})(D) \in S'] \leq e^{\epsilon} \cdot \Pr[(f \circ \mathcal{A})(D') \in S'] + \delta \]
This holds for any $S'$, proving that $f \circ \mathcal{A}$ is $(\epsilon, \delta)$-DP.

\textbf{Relevance to MLLM Fine-tuning:} In our context, the DP-SGD algorithm $\mathcal{A}$ takes the private dataset $\mathcal{D}_{\text{fine}}$ and produces the model parameters $\theta_{\text{fine}}$. The act of generating a prediction for a new input $x$, i.e., computing $M_{\theta_{\text{fine}}}(x)$, can be viewed as a post-processing function $f$ applied to $\theta_{\text{fine}}$. Therefore, the generated predictions inherit the same $(\epsilon, \delta)$-DP guarantee with respect to the fine-tuning dataset $\mathcal{D}_{\text{fine}}$.

\section{Motivation for Mechanism 1: Visual Token Pruning and Fusion}
\label{app:stage1_motivation}

This appendix details the motivation behind the visual input preprocessing performed in Mechanism 1 of our Dual-Priv Pruning method (Section~\ref{sec:methodology}). This stage operates \textit{before} the formal Differentially Private (DP) fine-tuning in Mechanism 2 (\S\ref{stage2:revised}) and is designed to address key challenges in applying DP to Multimodal Large Language Models (MLLMs). Specifically, it aims to reduce computation cost and potentially improve the utility outcome under DP constraints by modifying the visual token sequence.

\subsection{Addressing Computation Cost and Visual Redundancy}
Fine-tuning MLLMs using DP-SGD can be computationally demanding, due to the high number of visual tokens ($n$) generated by the vision encoder.  It has been observed that considerable redundancy exists within the visual tokens generated by Vision Transformers (ViTs), and not all tokens are equally important for downstream task performance~\cite{kong2022spvit, haurum2023tokens}. Building on the insight that attention scores often correlate with token importance~\cite{haurum2023tokens}, Mechanism 1 identifies and retains only the top-$K$ tokens receiving the highest aggregated attention from the [CLS] token. This selective pruning significantly shortens the sequence length processed in Mechanism 2, thereby directly reducing computation overhead. This strategy aligns with research exploring attention-based token pruning in ViTs~\cite{kong2022spvit, rao2021dynamicvit}.

\subsection{Preserving Context via Token Fusion}
While pruning reduces costs, simply discarding less attended tokens might remove valuable contextual information. To mitigate this, Mechanism 1 adopts a fusion strategy inspired by techniques that aim to compress information from pruned parts of a network or input~\cite{wei2023joint}. We merge the non-dominant tokens ($\mathcal{V}_{nd}$) into selected context tokens ($c$). This allows us to maintain a drastically reduced sequence length for efficiency while still incorporating a summarized representation of the less critical visual information, aiming for a better balance between computational savings and information preservation.
\subsection{Heuristic Noise Injection: Motivations and Potential Benefits}
The final step of Mechanism 1 introduces heuristic Gaussian noise to the fused context tokens ($c$) (Eq.~\eqref{fus_final}). This deliberate noise injection is multifaceted, aiming to potentially enhance the subsequent DP fine-tuning process:

\begin{itemize}
    \item \textbf{Regularization against DP Noise:} Adding noise is a known regularization technique~\citep{bishop1995training, noh2017regularizing}. Injecting noise specifically into the summarized, less critical token representation might act as \textbf{targeted input regularization}. This could potentially improve the model's robustness against the gradient perturbations inherent in the DP mechanism.

    \item \textbf{Encouraging Focus on Critical Tokens:} By introducing stochasticity primarily to the fused context token, the model might be implicitly encouraged during fine-tuning to rely more heavily on the stable, un-noised dominant tokens ($\mathcal{V}_d, v_{cls}$). This could help \textbf{preserve utility related to salient image features}.

    \item \textbf{Connection to Learning with Noise Priors:} Although mechanically different, this strategy shares a conceptual link with methods improving DP training by incorporating knowledge from noisy processes~\cite{tang2023differentially}.Our direct noise injection might serve a similar purpose by \textbf{preconditioning the model with input stochasticity}, potentially enhancing its resilience to the noise required for the DP guarantee in Mechanism 2.

    \item \textbf{Conceptual Input-Level Obfuscation:} While not contributing to the formal DP guarantee, manipulating the representation of less critical tokens with heuristic noise offers a degree of \textbf{data obfuscation at the input level}. This might provide some practical hardening against certain inference attacks targeting those specific, less informative image regions.
\end{itemize}

It is crucial to emphasize that the noise added in  Mechanism 2\ref{stage1} ($\sigma_{fuse}^2$) is \textbf{heuristic}. It is not calibrated according to DP principles and serves as a hyperparameter tuned for its potential benefits to utility and robustness.

\section{Motivation for Mechanism 2 Gradient-Update Pruning}
\label{app:stage2_motivation}

The post-noise adaptive update mechanism described in \cref{stage2:revised} is motivated by the goal of enhancing model utility under the constraints imposed by DP-SGD noise. Standard DP-SGD applies the noisy gradient $\tilde{g}$ (\cref{eq:unconditional_noise_main}) uniformly to all trainable parameters $\theta_{\text{train}}$. However, the added noise can significantly perturb or even dominate the true gradient signal, especially for parameter blocks where the original gradient magnitude was small or when operating under strict privacy budgets (requiring large $\sigma$). Applying updates based on such noise-dominated gradients might hinder convergence or lead to suboptimal performance.

Our strategy addresses this by analyzing the noisy gradient $\tilde{g}$ \textit{after} the privacy-preserving noise has been added. By partitioning $\tilde{g}$ into logical blocks $\tilde{g}_j$ and examining their $L_2$ norms $N_j = \|\tilde{g}_j\|_2$, we attempt to identify blocks where the signal likely outweighs the noise. The assumption is that a relatively large norm $N_j$ suggests that the original aggregated gradient component $\bar{\hat{g}}_j$ was sufficiently strong to persist despite the noise addition, thus indicating a more reliable update direction. Conversely, a small norm $N_j$ might indicate that the true signal was weak or was largely cancelled by the random noise vector.

The gating mechanism (\cref{eq:mask_generation_main_topk}) leverages this analysis. By generating a mask $M$ that selectively allows updates only for blocks with high noisy-gradient norms (i.e., where $M_j=1$), we filter out potentially detrimental updates arising from low-signal or noise-dominated gradient components. The final masked update (\cref{eq:masked_update_main}) focuses the optimization process on parameter blocks associated with stronger, potentially more informative, noisy gradient signals. This aims to improve the effective signal-to-noise ratio of the updates applied to the model, potentially leading to better convergence, improved utility, and a more favorable privacy-utility trade-off for the given privacy budget $(\epsilon, \delta)$.

\section{Baseline Details}
\label{app:baseline_details}

This section provides detailed descriptions, algorithms, and hyperparameter configurations for the baseline methods used in our comparative experiments.

\subsection{DP-SGD Baseline}
We implement the standard Differentially Private Stochastic Gradient Descent (DP-SGD) algorithm~\cite{abadi2016deep}, formally defined in {\ref{def:dpsgd}}. This method involves computing per-sample gradients, clipping their $L_2$ norms, averaging the clipped gradients, and adding calibrated Gaussian noise before updating the model parameters. It serves as the primary benchmark for differentially private optimization in deep learning. The hyperparameter configuration used for DP-SGD is detailed in Table~\ref{tab:dpsgd_params}.

\begin{table}[h] 
\centering
\caption{Hyperparameter Configuration for DP-SGD Baseline.}
\label{tab:dpsgd_params}
\begin{tabular}{@{}ll@{}}
\toprule
Parameter                 & Value \\ \midrule
Base Model                & LLAVA-7B~\cite{liu2023visual} \\
Fine-tuning Method        & LoRA~\cite{hu2022lora} \\
LoRA Rank ($r$)           & 128 \\
LoRA Alpha ($\alpha$)     & 256 \\
Optimizer                 & Adam~\cite{kingma2014adam} \\
Learning Rate             & 2e-4 \\
Batch Size                & 12 \\
Epochs                    & 1 \\ \midrule
\multicolumn{2}{@{}l}{\textbf{DP Parameters}} \\
Clipping Norm ($C$)       & 1.0 \\
Target $\delta$           & $\approx 1/N$ (Inverse dataset size) \\
Target $\epsilon$ Values  & \{1, 3, 8,$\infty$\} \\
Noise Multiplier ($\sigma$) & Calculated via RDP~\cite{DBLP:conf/csfw/Mironov17}  \\
                           & based on target $(\epsilon, \delta)$, $C$, $q$, and total steps. \\ \bottomrule
\end{tabular}
\end{table}
\subsection{DPZO Baseline}
DPZO (Differentially Private Zeroth-Order Optimization)~\cite{tang2024private} is a gradient-free DP optimization method. It approximates the gradient direction using finite differences based on random perturbations and privatizes only a scalar value representing the estimated directional derivative (loss difference). This avoids the memory overhead associated with storing per-sample gradients, but often requires significantly more iterations for convergence compared to DP-SGD. Algorithm~\ref{alg:dpzo_appendix} outlines the core mechanism adapted from~\cite{tang2024private}. The specific configuration used in our experiments is presented in Table~\ref{tab:dpzo_params_appendix}.
\begin{table}[h]
\centering
\caption{Hyperparameter Configuration for DPZO Baseline.}
\label{tab:dpzo_params_appendix} 
\scalebox{0.8}{\begin{tabular}{@{}ll@{}}
\toprule
Parameter                 & Value \\ \midrule
Base Model                & LLAVA-7B~\cite{liu2023visual} \\
Fine-tuning Method        & LoRA~\cite{hu2022lora} \\
LoRA Rank ($r$)           & 128 \\
LoRA Alpha ($\alpha$)     & 256 \\
Learning Rate ($\eta$)    & 2e-4  \\
Perturbation Scale ($\phi$) & 0.15 \\
Batch Size                & 12 \\
Epochs                    & 1 \\ \midrule
\multicolumn{2}{@{}l}{\textbf{DP Parameters}} \\
Clipping Norm ($C_{ZO}$)  & 1.0 \\
Target $\delta$           & $\approx 1/N$ \\
Target $\epsilon$ Values  & \{1, 3, 8, $\infty$\} \\
Noise Multiplier ($\sigma_{ZO}$) & Calculated via RDP accountant based on target $(\epsilon, \delta)$, $C_{ZO}$, $q=m/N$, $T$. \\ \bottomrule
\end{tabular}}
\end{table}

\section{Detailed Results on Medical Datasets}
\label{app:medical_results}

This section provides the detailed performance results for the experiments on the PathVQA and VQA-RAD datasets, as referenced in \cref{subsec:setup}. All experiments used a batch size (BS) of 12.
\begin{table}[!htbp] 
    \centering
    \caption{Detailed performance on PathVQA (BS=12). Higher is better for BLUE, EXT, F1. Best DP results in \textbf{bold}.}
    \label{tab:app_pathvqa_results} 
    \small 
    \setlength{\tabcolsep}{3.5pt} 
    \sisetup{round-mode=places, round-precision=4, table-format=1.4} 
    \begin{tabular}{@{}l | SSS | SSS | SSS @{}}
        \toprule
        \multirow{2}{*}{$\epsilon$} & \multicolumn{3}{c|}{Ours (Dual-Priv)} & \multicolumn{3}{c|}{DPZO} & \multicolumn{3}{c@{}}{DP-SGD} \\
        \cmidrule(lr){2-4} \cmidrule(lr){5-7} \cmidrule(lr){8-10}
        & {BLUE} & {EXT} & {F1} & {BLUE} & {EXT} & {F1} & {BLUE} & {EXT} & {F1} \\
        \midrule
        1    & \bfseries \textbf{0.7385} & \bfseries 0.3840 & \bfseries 0.3792 & 0.6534 & 0.0301 & 0.0592 & 0.7222 & 0.3732 & 0.3675 \\
        3    & \bfseries \textbf{0.7263} & \bfseries 0.3738 & \bfseries 0.3701 & 0.6534 & 0.0301 & 0.0592 & 0.7257 & 0.3712 & 0.3653 \\
        8    & \bfseries \textbf{0.7195} & \bfseries 0.3763 & \bfseries 0.3713 & 0.6534 & 0.0301 & 0.0592 & 0.7140 & 0.3683 & 0.3635 \\
        $\infty$ & 0.7430 & 0.3880 & 0.3841 & 0.6534 & 0.0301 & 0.0592 & 0.7182 & \bfseries 0.3927 & \bfseries 0.3879 \\ 
        \bottomrule
    \end{tabular}
\end{table}

\begin{table}[!htbp] 
    \centering
    \caption{Detailed accuracy (\%) on VQA-RAD (BS=12). Higher is better. Best DP result in \textbf{bold}.}
    \label{tab:app_vqarad_results} 
    \small 
    \setlength{\tabcolsep}{5pt} 
    \sisetup{round-mode=places, round-precision=1, table-format=2.1} 
    \begin{tabular}{@{}l S S S@{}}
        \toprule
        $\epsilon$ & {Ours (Dual-Priv)} & {DPZO} & {DP-SGD} \\
        \midrule
        1    & \bfseries \textbf{48.6} & 0.0 & 47.3 \\
        3    & \bfseries \textbf{48.8} & 0.0 & 48.1 \\
        8    & \bfseries \textbf{49.0} & 0.0 & 46.8 \\
        $\infty$ & 47.9 & 0.0 & \bfseries 48.3 \\ 
        \bottomrule
    \end{tabular}
\end{table}
\section{Additional Details on Membership Inference Attack}
\label{app:mia_results}
This section provides the additional details for the experiments with membership inference attack, as referenced in \cref{subsec:mia_results}. All experiments used a batch size(BS) of 12. We randomly sample 6,000 image-text pairs from the ROCOV2 dataset for evaluation and randomly sampled 3000 image-text pairs as the member dataset for training. To fit the LLaVA-VQA formulation, we randomly use these prompts:"Please describe the image in detail.",
    "What is shown in this medical image?",
    "Describe the contents of this image.",
    "What does this image depict?",
    "Provide a detailed description of this image.",
    "Please analyze this medical image.",
    "Describe the medical image in detail.",
    "Describe the condition depicted in the image.",
    "Please provide a caption for this image."
\section{Limitations}
\label{app:limitations}
Our study demonstrates the effectiveness of Dual-Priv Pruning for DP fine-tuning MLLMs. While our evaluations on a 7B MLLM are thorough, extending the assessment to MLLMs of substantially different scales would provide a broader understanding of the approach's scalability. 

\section{Broader Impacts}
\label{app:broader_impacts}

The development of Dual-Priv Pruning contributes to the critical area of privacy-preserving machine learning, particularly for Multimodal Large Language Models (MLLMs). The primary societal benefit lies in its potential to significantly enhance data privacy when fine-tuning MLLMs on sensitive datasets. By integrating Differential Privacy (DP) with improved efficiency and utility, our work can empower the responsible use of MLLMs in domains handling personal information, such as healthcare or finance, thereby protecting individuals from data leakage. This advancement may also lower barriers to adopting privacy-enhancing technologies, encouraging a broader shift towards responsible AI practices and facilitating research on valuable sensitive datasets that might otherwise remain underutilized due to privacy risks. Ultimately, robust privacy measures like those explored can foster greater public trust in AI systems, which is vital for their ethical and successful integration into society.

However, it is important to consider the broader context. While DP offers strong mathematical privacy guarantees, these are contingent upon correct implementation and careful parameter selection, and they address specific threats related to individual data contributions rather than all conceivable privacy concerns. A nuanced understanding is crucial to avoid a false sense of absolute security. The inherent trade-off between privacy protection and model utility, though mitigated by our approach, persists; in certain high-stakes applications, even minor performance degradation due to DP noise could have notable implications if not carefully weighed. Furthermore, the expertise required to effectively implement and tune DP mechanisms remains a consideration for broader accessibility. While our method focuses on the privacy of training data, the underlying MLLM technology itself, regardless of how it's fine-tuned, could still be subject to misuse if its outputs are leveraged for unintended or harmful purposes.

Our research is a step towards more responsible AI development. We believe continued efforts in the community are essential to further refine the balance between privacy and utility, enhance the usability of DP tools, and promote comprehensive education on both the capabilities and limitations of such privacy-enhancing technologies. Addressing fairness and bias within DP-trained models also remains an important ongoing pursuit. This work is presented as foundational research to advance privacy in MLLM fine-tuning, with the anticipation that its net impact will be positive by enabling more secure and trustworthy AI applications.
\section{Algorithm for Baselines and Dual-priv Pruning}
\label{app:algo_details}
\cref{alg:dp_adaptive_update_appendix} provides the detailed step-by-step procedure for the Stage 2 DP fine-tuning process described in Section~4.2 of the main paper. While  
\cref{alg:dpsgd_appendix} outlines the standard DP-SGD baseline, and \cref{alg:dpzo_appendix} details the DPZO baseline.

\begin{algorithm}[htbp]
\caption{Dual-Priv Pruning: Mechanism 2 (DP Fine-tuning with Gradient-Update Pruning)}
\label{alg:dp_adaptive_update_appendix}
\begin{algorithmic}[1]
\Require Initial trainable parameters $\theta_{\text{train}_0}$, dataset $D = \{(\mathcal{V}'_i, \mathcal{T}_i)\}_{i=1}^N$ (with pre-processed visual inputs $\mathcal{V}'_i$), learning rate schedule $\eta_t$, gradient clipping norm $C$, noise multiplier $\sigma$ (derived from target $\epsilon, \delta$), batch size $m$, total training steps $T$, number of logical parameter blocks $J$ in $\theta_{\text{train}}$, top-K percentage $P_K$.
\For{$t = 1, \dots, T$}
    \State Sample minibatch $\xi_t = \{(\mathcal{V}'_k, \mathcal{T}_k)\}_{k=1}^m \subset D$ of size $m$.
    \State Initialize list of per-sample gradients $G_{list} = []$.
    \For{each sample $(\mathcal{V}'_k, \mathcal{T}_k) \in \xi_t$}
        \State Compute gradient $g_k \leftarrow \nabla_{\theta_{\text{train}_{t-1}}} \mathcal{L}(\theta_{\text{train}_{t-1}}; (\mathcal{V}'_k, \mathcal{T}_k))$.
        \State Clip gradient: $\hat{g}_k \leftarrow g_k / \max(1, \|g_k\|_2 / C)$.
        \State Append $\hat{g}_k$ to $G_{list}$.
    \EndFor
    \State Aggregate clipped gradients: $\bar{\hat{g}} \leftarrow \frac{1}{m} \sum_{\hat{g}_k \in G_{list}} \hat{g}_k$.
    \State Add Gaussian noise: $\tilde{g} \leftarrow \bar{\hat{g}} + \mathcal{N}\left(0, \frac{\sigma^2 C^2}{m^2} I_{d_{\text{train}}}\right)$.
    \State Partition $\tilde{g}$ into $J$ components $\{\tilde{g}_1, \dots, \tilde{g}_J\}$ corresponding to logical parameter blocks.
    \State Compute $L_2$ norms for each block: $N_j \leftarrow \|\tilde{g}_j\|_2$ for $j=1, \dots, J$.
    \State Initialize mask $M$ as a zero tensor with the same block structure as $\theta_{\text{train}}$.
    \State Let $K_{\text{count}} \leftarrow \lceil (P_K/100) \cdot J \rceil$.
    \State Let $\mathcal{S}_{\text{top\_indices}}$ be the set of indices of the $K_{\text{count}}$ blocks with the largest norms $N_j$.
    \For{each block index $j \in \mathcal{S}_{\text{top\_indices}}$}
        \State Set corresponding part of mask $M_j \leftarrow \mathbf{1}$ (vector/matrix of ones for block $j$).
    \EndFor
    \State Update parameters: $\theta_{\text{train}_t} \leftarrow \theta_{\text{train}_{t-1}} - \eta_t \cdot (M \odot \tilde{g})$.
\EndFor
\State \Return $\theta_{\text{train}_T}$.
\end{algorithmic}
\end{algorithm}

\begin{algorithm}[htbp]
\caption{Differentially Private Stochastic Gradient Descent (DP-SGD, adapted from \cite{abadi2016deep})}
\label{alg:dpsgd_appendix}
\begin{algorithmic}[1]
\Require Initial model parameters $\theta_0$, dataset $D = \{(\mathcal{I}_i, \mathcal{T}_i)\}_{i=1}^N$ (or generic $x_i$), learning rate $\eta_t$, clipping norm $C$, noise multiplier $\sigma$, batch size $m$, total steps $T$.
\For{$t = 1, \dots, T$}
    \State Sample minibatch $\xi_t = \{x_k\}_{k=1}^m \subset D$ of size $m$.
    \State Initialize list of per-sample gradients $G_{list} = []$.
    \For{each sample $x_k \in \xi_t$}
        \State Compute gradient $g_k \leftarrow \nabla_{\theta_{t-1}} \mathcal{L}(\theta_{t-1}; x_k)$.
        \State Clip gradient: $\hat{g}_k \leftarrow g_k / \max(1, \|g_k\|_2 / C)$.
        \State Append $\hat{g}_k$ to $G_{list}$.
    \EndFor
    \State Aggregate clipped gradients: $\bar{\hat{g}} \leftarrow \frac{1}{m} \sum_{\hat{g}_k \in G_{list}} \hat{g}_k$.
    \State Add Gaussian noise: $\tilde{g} \leftarrow \bar{\hat{g}} + \mathcal{N}\left(0, \frac{\sigma^2 C^2}{m^2} I_d\right)$.
    \State Update parameters: $\theta_t \leftarrow \theta_{t-1} - \eta_t \cdot \tilde{g}$.
\EndFor
\State \Return $\theta_T$.
\end{algorithmic}
\end{algorithm}

\begin{algorithm}[htbp]
\caption{DPZO Core Mechanism (Simplified, adapted from \cite{tang2024private})}
\label{alg:dpzo_appendix} 
\begin{algorithmic}[1]
\Require Model parameters $\theta$, dataset $D$, learning rate $\eta$, perturbation scale $\phi$, clipping threshold $C_{ZO}$, noise scale $\sigma_{ZO}$, batch size $m$, total steps $T$.
\For{$t = 1, \dots, T$}
    \State Sample batch $B \subset D$.
    \State Sample random direction $z_t \sim \mathcal{N}(0, I_d)$.
    \State Set $\theta^+ \leftarrow \theta_{t-1} + \phi z_t$, $\theta^- \leftarrow \theta_{t-1} - \phi z_t$.
    \State Initialize loss differences list $L_{\text{diff}} = []$.
    \For{each sample $(\mathcal{I}_i, \mathcal{T}_i) \in B$}
        \State Compute $l_i = \mathcal{L}(\theta^+; (\mathcal{I}_i, \mathcal{T}_i)) - \mathcal{L}(\theta^-; (\mathcal{I}_i, \mathcal{T}_i))$.
        \State Clip difference: $\hat{l}_i \leftarrow \max(-C_{ZO}, \min(l_i, C_{ZO}))$.
        \State Append $\hat{l}_i$ to $L_{\text{diff}}$.
    \EndFor
    \State Aggregate clipped differences: $\bar{l} \leftarrow \frac{1}{|B|} \sum_{\hat{l}_i \in L_{\text{diff}}} \hat{l}_i$.
    \State Add noise to privatize the average difference: $s \leftarrow \bar{l} + \mathcal{N}(0, \sigma_{ZO}^2 C_{ZO}^2 / |B|^2)$.
    \State Update parameters: $\theta_t \leftarrow \theta_{t-1} - \eta \cdot s \cdot z_t / (2\phi)$.
\EndFor
\State \Return $\theta_T$.
\end{algorithmic}
\end{algorithm}

\clearpage
\section*{NeurIPS Paper Checklist}
\begin{enumerate}

\item {\bf Claims}
    \item[] Question: Do the main claims made in the abstract and introduction accurately reflect the paper's contributions and scope?
    \item[] Answer: \answerYes{} 
    \item[] Justification: The abstract and introduction accurately reflect the paper’s contributions.

\item {\bf Limitations}
    \item[] Question: Does the paper discuss the limitations of the work performed by the authors?
    \item[] Answer: \answerYes{} 
    \item[] Justification: See Appendix \ref{app:limitations}

\item {\bf Theory assumptions and proofs}
    \item[] Question: For each theoretical result, does the paper provide the full set of assumptions and a complete (and correct) proof?
    \item[] Answer: \answerYes{} 
    \item[] Justification: See \cref{sec:preliminaries}, \cref{app:dp_facts}, \cref{app:post_processing} and \cref{app:rdp_details}

    \item {\bf Experimental result reproducibility}
    \item[] Question: Does the paper fully disclose all the information needed to reproduce the main experimental results of the paper to the extent that it affects the main claims and/or conclusions of the paper (regardless of whether the code and data are provided or not)?
    \item[] Answer: \answerYes{}  
    \item[] Justification: We claim the details of methods and the experiments settings in our paper.

\item {\bf Open access to data and code}
    \item[] Question: Does the paper provide open access to the data and code, with sufficient instructions to faithfully reproduce the main experimental results, as described in supplemental material?
    \item[] Answer: \answerYes{} 
    \item[] Justification: We include the code in our supplemental material.

\item {\bf Experimental setting/details}
    \item[] Question: Does the paper specify all the training and test details (e.g., data splits, hyperparameters, how they were chosen, type of optimizer, etc.) necessary to understand the results?
    \item[] Answer: \answerYes{}  
    \item[] Justification:: The experimental settings are detailed in \cref{subsec:setup}

\item {\bf Experiment statistical significance}
    \item[] Question: Does the paper report error bars suitably and correctly defined or other appropriate information about the statistical significance of the experiments?
    \item[] Answer: \answerNo{} 
    \item[] Justification: We believe the pattern is clear.

\item {\bf Experiments compute resources}
    \item[] Question: For each experiment, does the paper provide sufficient information on the computer resources (type of compute workers, memory, time of execution) needed to reproduce the experiments?
    \item[] Answer: \answerYes{}{} 
    \item[] Justification:See \cref{sec:experiments}
    
\item {\bf Code of ethics}
    \item[] Question: Does the research conducted in the paper conform, in every respect, with the NeurIPS Code of Ethics \url{https://neurips.cc/public/EthicsGuidelines}?
    \item[] Answer: \answerYes{} 
    \item[] Justification: It does.

\item {\bf Broader impacts}
    \item[] Question: Does the paper discuss both potential positive societal impacts and negative societal impacts of the work performed?
    \item[] Answer: \answerYes{} 
    \item[] Justification: See \cref{app:broader_impacts}
    
\item {\bf Safeguards}
    \item[] Question: Does the paper describe safeguards that have been put in place for responsible release of data or models that have a high risk for misuse (e.g., pretrained language models, image generators, or scraped datasets)?
    \item[] Answer: \answerNA{} 
    \item[] Justification: The paper poses no such risks.

\item {\bf Licenses for existing assets}
    \item[] Question: Are the creators or original owners of assets (e.g., code, data, models), used in the paper, properly credited and are the license and terms of use explicitly mentioned and properly respected?
    \item[] Answer: \answerYes{} 
    \item[] Justification: We cite all works properly.

\item {\bf New assets}
    \item[] Question: Are new assets introduced in the paper well documented and is the documentation provided alongside the assets?
    \item[] Answer: \answerYes{} 
    \item[] Justification: We offer documentation alongside our code. The anonymized repository including code and documentation can be found at: \url{https://anonymous.4open.science/r/Dual-priv-pruning-AE7E}

\item {\bf Crowdsourcing and research with human subjects}
    \item[] Question: For crowdsourcing experiments and research with human subjects, does the paper include the full text of instructions given to participants and screenshots, if applicable, as well as details about compensation (if any)? 
    \item[] Answer: \answerNA{} 
    \item[] Justification:: The paper does not involve crowdsourcing nor research with human subjects.

\item {\bf Institutional review board (IRB) approvals or equivalent for research with human subjects}
    \item[] Question: Does the paper describe potential risks incurred by study participants, whether such risks were disclosed to the subjects, and whether Institutional Review Board (IRB) approvals (or an equivalent approval/review based on the requirements of your country or institution) were obtained?
    \item[] Answer: \answerYes{} 
    \item[] Justification: The paper does not involve crowdsourcing nor research with human subjects.

\item {\bf Declaration of LLM usage}
    \item[] Question: Does the paper describe the usage of LLMs if it is an important, original, or non-standard component of the core methods in this research? Note that if the LLM is used only for writing, editing, or formatting purposes and does not impact the core methodology, scientific rigorousness, or originality of the research, declaration is not required.
    \item[] Answer: \answerYes{} 
    \item[] Justification: The core methodology of this research is centered on the differential private fine-tuning of Multimodal Large Language Models (MLLMs).

\end{enumerate}

\end{document}